\documentclass[twocolumn]{aastex62}
\usepackage[utf8]{inputenc}
\usepackage [autostyle, english = american]{csquotes}
\usepackage{enumitem}
\usepackage{textcomp}
\usepackage{chngpage}
\usepackage{amsmath}
\usepackage{footmisc}
\usepackage{float}
\usepackage{bm}
\usepackage{mathtools, nccmath}
\MakeOuterQuote{"}
\usepackage{xspace}
\usepackage{tablefootnote}
\usepackage{threeparttable}
\newcommand{\Kepler}{\emph{Kepler}}
\newcommand{\TESS}{\emph{TESS}}
\newcommand{\Gaia}{\emph{Gaia}}
\newcommand{\SOPHIE}{\emph{SOPHIE}}
\let\oldAA\AA
\renewcommand{\AA}{\text{\normalfont\oldAA}}
\newcommand{\kms}{km\,s$^{-1}$}

\newcommand{\masy}{mas\,y$^{-1}$}

\newcommand{\Msun}{\mbox{$M_{\odot}$}}
\newcommand{\Rsun}{\mbox{$R_{\odot}$}}
\newcommand{\rearth}{R$_{\oplus}$}
\newcommand{\gccc}{g\,cm$^{-3}$}

\newcommand{\teff}{$T_{\rm eff}$}
\newcommand{\logg}{$\log g$}

\newcommand{\allesfitter}{\texttt{allesfitter}} 
\newcommand{\isoclassify}{\texttt{isoclassify}}
\newcommand{\isochrones}{\texttt{isochrones}}
\newcommand{\VESPA}{\texttt{vespa}}
\newcommand{\lightkurve}{\texttt{lightkurve}}
\newcommand{\forecaster}{\texttt{forecaster}}

\newcommand\clearrow{\global\let\rowmac\relax}

\submitjournal{AAS Journals}
\received{February 10, 2020}
\revised{June 19, 2020}
\accepted{June 25, 2020}

\shorttitle{HD~191939: Three Sub-Neptunes Transiting a Sun-like Star Only 54 pc Away}
\shortauthors{Badenas-Agusti et al.}
\graphicspath{{./}{figures/}}
\begin{document}

\title{HD~191939: Three Sub-Neptunes Transiting a Sun-like Star Only 54 pc Away}

\author[0000-0003-4903-567X]{Mariona Badenas-Agusti}
\affil{Department of Earth, Atmospheric and Planetary Sciences, Massachusetts Institute of Technology, Cambridge, MA 02139, USA}
\affil{Department of Physics, and Kavli Institute for Astrophysics and Space Research, Massachusetts Institute of Technology, Cambridge, MA 02139, USA}

\author[0000-0002-3164-9086]{Maximilian N.\ G{\"u}nther}
\affil{Department of Physics, and Kavli Institute for Astrophysics and Space Research, Massachusetts Institute of Technology, Cambridge, MA 02139, USA}
\affil{Juan Carlos Torres Fellow}

\author[0000-0002-6939-9211]{Tansu Daylan}
\affil{Department of Physics, and Kavli Institute for Astrophysics and Space Research, Massachusetts Institute of Technology, Cambridge, MA 02139, USA}
\affil{Kavli Fellow}

\author[0000-0001-5442-1300]{Thomas Mikal-Evans}
\affil{Department of Physics, and Kavli Institute for Astrophysics and Space Research, Massachusetts Institute of Technology, Cambridge, MA 02139, USA}

\author[0000-0001-7246-5438]{Andrew Vanderburg}
\affil{Department of Astronomy, The University of Texas at Austin, Austin, TX 78712, USA}
\affil{NASA Sagan Fellow}

\author[0000-0003-0918-7484]{Chelsea X.\ Huang}
\affil{Department of Physics, and Kavli Institute for Astrophysics and Space Research, Massachusetts Institute of Technology, Cambridge, MA 02139, USA}
\affil{Juan Carlos Torres Fellow}

\author[0000-0003-0593-1560]{Elisabeth Matthews}
\affil{Department of Physics, and Kavli Institute for Astrophysics and Space Research, Massachusetts Institute of Technology, Cambridge, MA 02139, USA}

\author[0000-0002-3627-1676]{Benjamin V.\ Rackham}
\affil{Department of Earth and Planetary Sciences, MIT, 77 Massachusetts Avenue, Cambridge, MA 02139, USA}
\affil{Department of Physics, and Kavli Institute for Astrophysics and Space Research, Massachusetts Institute of Technology, Cambridge, MA 02139, USA}
\affil{51 Pegasi b Fellow}

\author[0000-0001-6637-5401]{Allyson Bieryla}
\affil{Harvard-Smithsonian | Center for Astrophysics, 60 Garden St, Cambridge, MA 02138, USA}

\author[0000-0002-3481-9052]{Keivan G.\ Stassun}
\affil{Department of Physics and Astronomy, Vanderbilt University, 6301 Stevenson Center Ln., Nashville, TN 37235, USA}
\affil{Department of Physics, Fisk University, 1000 17th Ave. N., Nashville, TN 37208, USA}

\author{Stephen R.\ Kane}
\affil{Department of Earth and Planetary Sciences, University of California, Riverside, CA 92521, USA}

\author{Avi Shporer}
\affil{Department of Physics, and Kavli Institute for Astrophysics and Space Research, Massachusetts Institute of Technology, Cambridge, MA 02139, USA}

\author[0000-0003-3504-5316]{Benjamin J.\ Fulton}
\affil{California Institute of Technology, Pasadena, CA 91125, USA}
\affil{IPAC-NASA Exoplanet Science Institute Pasadena, CA 91125, USA}

\author[0000-0002-0139-4756]{Michelle L. Hill}
\affil{Department of Earth and Planetary Sciences, University of California, Riverside, CA 92521, USA}

% CARMENES Consortium representatives, in alphabetical order

\author[0000-0002-7031-7754]{Grzegorz Nowak}
\affil{Instituto de Astrofísica de Canarias (IAC), E-38200 La Laguna, Tenerife, Spain}
\affil{Departamento de Astrofísica, Universidad de La Laguna, E-38206 La Laguna, Tenerife, Spain}

\author[0000-0002-6689-0312]{Ignasi Ribas}
\affil{Institut de Ciències de l'Espai (ICE, CSIC), Campus UAB, C/Can Magrans, s/n, 08193 Bellaterra, Spain}
\affil{Institut d'Estudis Espacials de Catalunya (IEEC), 08034 Barcelona, Spain}

\author[0000-0003-0987-1593]{Enric Pallé}
\affil{Instituto de Astrofísica de Canarias (IAC), E-38200 La Laguna, Tenerife, Spain}
\affil{Departamento de Astrofísica, Universidad de La Laguna, E-38206 La Laguna, Tenerife, Spain}

% #### TESS Architects in alphabetical order

\author{Jon M.\ Jenkins}
\affil{NASA Ames Research Center, Moffett Field, CA, 94035, USA}

\author[0000-0001-9911-7388]{David W.\ Latham}
\affil{Harvard-Smithsonian | Center for Astrophysics, 60 Garden St, Cambridge, MA 02138, USA}

\author[0000-0002-6892-6948]{Sara Seager}
\affil{Department of Physics, and Kavli Institute for Astrophysics and Space Research, Massachusetts Institute of Technology, Cambridge, MA 02139, USA}
\affil{Department of Earth, Atmospheric and Planetary Sciences, Massachusetts Institute of Technology, Cambridge, MA 02139, USA}
\affil{Department of Aeronautics and Astronautics, Massachusetts Institute of Technology, Cambridge, MA 02139, USA}

\author{George R.\ Ricker}
\affil{Department of Physics, and Kavli Institute for Astrophysics and Space Research, Massachusetts Institute of Technology, Cambridge, MA 02139, USA}

\author{Roland K.\ Vanderspek}
\affil{Department of Physics, and Kavli Institute for Astrophysics and Space Research, Massachusetts Institute of Technology, Cambridge, MA 02139, USA}

\author[0000-0002-4265-047X]{Joshua N.\ Winn}
\affil{Department of Astrophysical Sciences, Princeton University, 4 Ivy Lane, Princeton, NJ 08544, USA}

% #### follow-up efforts +  (alphabetically)

\author[0000-0002-1847-9481]{Oriol Abril-Pla}
\affil{Statistics Division, Universitat Pompeu Fabra, Ramon Trias Fargas 25-27, 08005 Barcelona, Spain} 

\author[0000-0001-6588-9574]{Karen A.\, Collins}
\affil{Harvard-Smithsonian | Center for Astrophysics, 60 Garden St, Cambridge, MA 02138, USA}

\author[0000-0002-4308-2339]{Pere Guerra Serra}
\affil{Observatori Astronòmic Albanyà, Camí de Bassegoda s/n, E-17733, Albanyà, Spain} 

\author[0000-0002-8052-3893]{Prajwal Niraula}
\affil{Department of Earth, Atmospheric and Planetary Sciences, Massachusetts Institute of Technology, Cambridge, MA 02139, USA}

\author[0000-0003-4408-0463]{Zafar Rustamkulov}
\affil{Department of Earth and Planetary Sciences, Johns Hopkins University, Baltimore, MD, USA}

% #### (TESS) Contributing authors 

\author[0000-0001-7139-2724]{Thomas Barclay}
\affil{NASA Goddard Space Flight Center, Greenbelt, MD 20771, USA}
\affil{University of Maryland, Baltimore County, 1000 Hilltop Cir, Baltimore, MD 21250, USA}

\author{Ian J.\ M.\ Crossfield}
\affil{Department of Physics and Astronomy, The University of Kansas, 1251 Wescoe Hall Drive, Lawrence, KS, 66045, USA}
\affil{Department of Physics, and Kavli Institute for Astrophysics and Space Research, Massachusetts Institute of Technology, Cambridge, MA 02139, USA}

\author[0000-0002-2532-2853]{Steve B.\ Howell}
\affil{NASA Ames Research Center, Moffett Field, CA, 94035, USA}

\author[0000-0002-5741-3047]{David R.\ Ciardi}
\affil{Caltech/IPAC, 1200 E. California Blvd. Pasadena, CA 91125, USA}

\author{Erica J.\ Gonzales}
\affil{Department of Astronomy and Astrophysics, University of California Santa Cruz, 1156 High St, Santa Cruz, CA 95060, USA}
\affil{NSF Graduate Research Fellowship Program Fellow}

\author[0000-0001-5347-7062]{Joshua E. Schlieder}
\affil{NASA Goddard Space Flight Center, Greenbelt, MD 20771, USA}

\author[0000-0003-1963-9616]{Douglas A.\ Caldwell}
\affil{SETI Institute/NASA Ames Research Center, Moffett Field, CA 94035, USA}

\author[0000-0002-9113-7162]{Michael Fausnaugh}
\affil{Department of Physics, and Kavli Institute for Astrophysics and Space Research, Massachusetts Institute of Technology, Cambridge, MA 02139, USA}

\author{Scott McDermott}
\affil{Proto-Logic LLC, 1718 Euclid Street NW, Washington, DC 20009, USA}

\author[0000-0001-8120-7457]{Martin Paegert}
\affil{Harvard-Smithsonian | Center for Astrophysics, 60 Garden St, Cambridge, MA 02138, USA}

\author[0000-0002-3827-8417]{Joshua Pepper}
\affiliation{Department of Physics, Lehigh University, 16 Memorial Drive East, Bethlehem, PA 18015, USA}

\author{Mark E.\ Rose}
\affil{NASA Ames Research Center, Moffett Field, CA, 94035}

\author{Joseph D.\ Twicken}
\affil{SETI Institute/NASA Ames Research Center, Moffett Field, CA 94035\hspace*{24pt}}

\correspondingauthor{Mariona Badenas-Agusti}
\email{mbadenas@mit.edu}

\begin{abstract}

We present the discovery of three sub-Neptune-sized planets transiting the nearby and bright Sun-like star HD~191939 (TIC~269701147, TOI~1339), a $K_{s}=7.18$\,magnitude G8 V dwarf at a distance of only 54\,parsecs. We validate the planetary nature of the transit signals by combining five months of data from the \textit{Transiting Exoplanet Survey Satellite} with follow-up ground-based photometry, archival optical images, radial velocities, and high angular resolution observations. The three sub-Neptunes have similar radii ($R_{b} = 3.42^{+0.11}_{-0.11}\,R_{\oplus}$, $R_{c}=3.23_{-0.11}^{+0.11}\,R_{\oplus}$, and $R_{d}=3.16_{-0.11}^{+0.11}\,R_{\oplus}$) and their orbits are consistent with a stable, circular, and co-planar architecture near mean motion resonances of 1:3 and 3:4 ($P_{b}=8.88$\,days, $P_{c}=28.58$\,days, and $P_{d}=38.35$\,days). The HD~191939 system is an excellent candidate for precise mass determinations of the planets with high-resolution spectroscopy due to the host star's brightness and low chromospheric activity. 
Moreover, the system's compact and near-resonant nature can provide an independent way to measure planetary masses via transit timing variations while also enabling dynamical and evolutionary studies. Finally, as a promising target for multi-wavelength transmission spectroscopy of all three planets' atmospheres, HD~191939 can offer valuable insight into multiple sub-Neptunes born from a proto-planetary disk that may have resembled that of the early Sun.

\end{abstract}

.

\section{Introduction} \label{sec:intro}

The \textit{Transiting Exoplanet Survey Satellite} (\TESS{}, \citealt{Ricker_2014}) was designed to detect transiting "super-Earths" ($R_{p}=1.25-2\, R_{\oplus}$, $M_{p} \approx 1-10\, M_{\oplus}$) and "sub-Neptunes" ($R_{p}=2-4\, R_{\oplus}$, $M_{p} \approx 10-40\,M_{\oplus}$) around the nearest and brightest main-sequence stars. As a result, planets detected by \TESS{} will be some of the best candidates for follow-up spectroscopy and future atmospheric characterization studies.  

\begin{figure*}[ht]
    \centering
    \includegraphics[width=\textwidth]{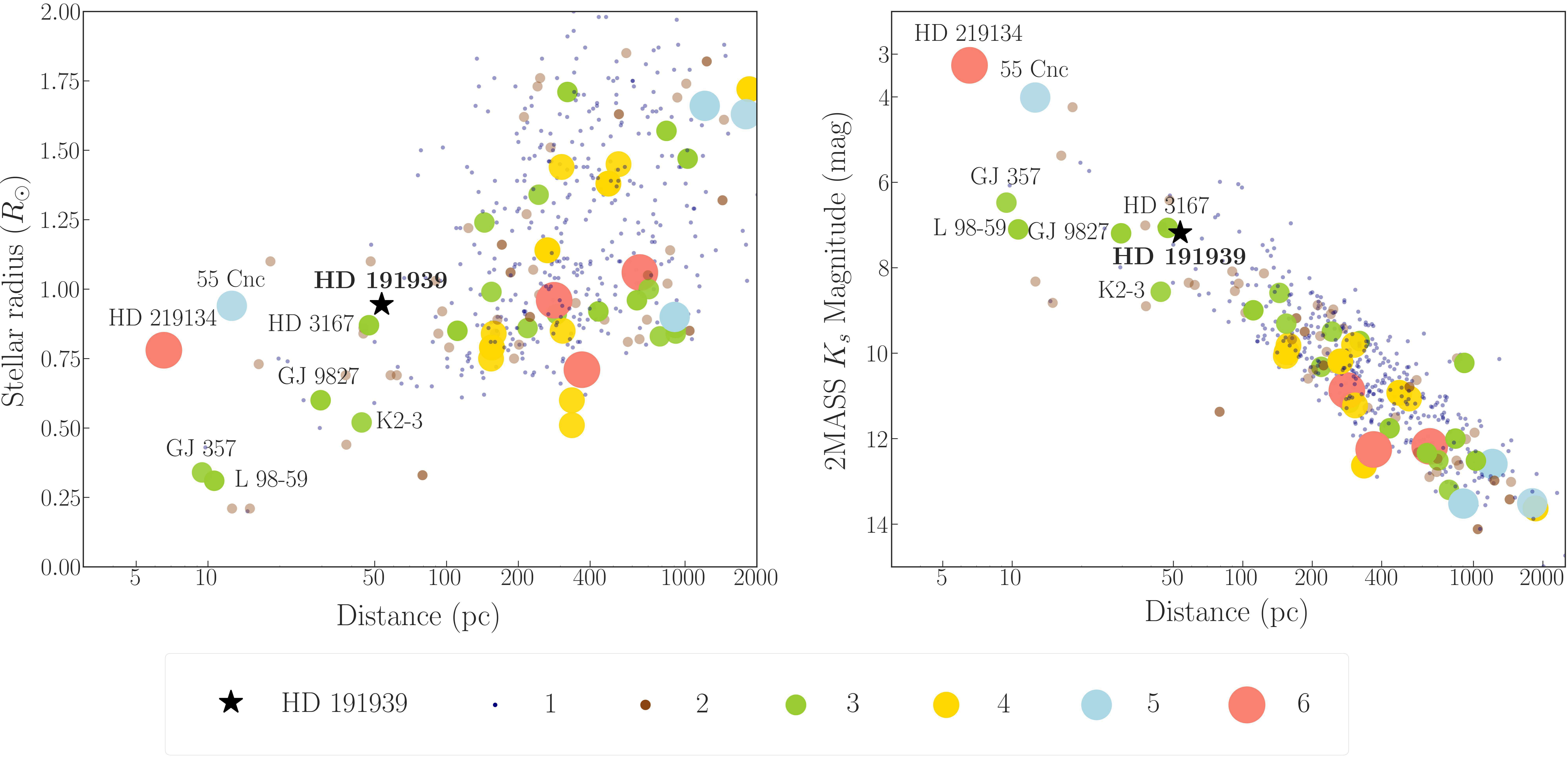}
    \caption{Host star radius and $K_{s}$ magnitude of confirmed single- and multi-planetary systems. In these views, we only show systems with measured masses and a relative error in planet mass, planet radius, and host star radius less than $30\%$. Names are only displayed for stars with at least three planets and with distances less than 100\,pc. The size and color of each system depends on the number of planets that it hosts. Data were retrieved from the NASA Exoplanet Archive in May 2020.}
    \label{fig:rs_kmag_vs_dist}
\end{figure*}

Since the beginning of science operations in 2018, \TESS{} has discovered about a dozen multi-transiting planet systems (e.g. \citealt{Huang:2018, Dragomir2019, Guenther2019, Quinn2019}), including some of the brightest known to date, 
thus yielding prime targets for detailed characterization \citep[e.g.][]{Huang:2018, Dragomir2019, Guenther2019, Quinn2019}.  These "multis" are excellent laboratories to perform comparative exoplanetology and learn about planetary formation and evolutionary processes in the controlled environment of the host star. Moreover, they often have a greater scientific potential than single-planet systems because they can be characterized comprehensively beyond the conventional methods of transit photometry and radial velocity (RV) observations \citep{ragozzine:2010}. For example, measurements of transit timing variations (TTVs) can help constrain planetary masses and orbital architectures (\citealt{Miralda:2002}, \citealt{Agol:2005}). For multis amenable to atmospheric characterization, transmission spectroscopy can shed light on the shared properties of planets born from the same proto-planetary disk.

The \Kepler{} mission revealed that multi-transiting planet systems are ubiquitous \citep{Latham:2011, lissauer2011, Lissauer:2014, Rowe:2014}, particularly in the super-Earth to mini-Neptune regime (e.g. \citealt{Howard:2010}; \citealt{Fressin:2013}). Despite their widespread occurrence, the majority of \Kepler{} multis are too faint, distant, and/or small to precisely determine planetary masses with independent (RV) surveys. Consequently, many of them lack mass and density measurements. With \TESS, however, the population of multis amenable to follow-up studies will grow as nearer and brighter systems are detected. Many of these \TESS{} discoveries will be sub-Jovian-sized planets well suited for spectroscopic studies of planetary masses (e.g. \citealt{Cloutier:2018}) and atmospheres (e.g. \citealt{Kempton:2018}) due to their larger sizes and their host star's proximity and brightness. 

Here we focus on HD~191939 (TOI~1339, \linebreak TIC~269701147), a bright ($V=8.97$\,mag, $K=7.18$\,mag), nearby ($d=53.48^{+0.19}_{-0.20}$\,pc), Sun-like (G8 V) star with a radius of $R_{*}=0.945\pm0.021\,R_{\odot}$, a mass of $M_{*}=0.92\pm0.06\,M_{\odot}$, and a temperature of $T_\mathrm{eff}=5400\pm50$\,K. Using \TESS{} data from sectors 15--19, we present the discovery of three sub-Neptune-sized planets around HD~191939 and validate their transit signals with archival optical images, RVs, ground-based photometric follow-up, and high-resolution imaging. At a distance of only $54$\,pc, HD~191939 is one of the nearest and brightest multi-transiting planet systems known to date (see \autoref{fig:rs_kmag_vs_dist}). Due to the host star's proximity, brightness, and low chromospheric activity, this multi is an excellent target for follow-up photometric and spectroscopic studies. As we step into the era of the \textit{James Webb Space Telescope} (JWST), HD~191939 is a promising candidate for detailed atmospheric characterization as well. 

This paper is organized as follows. Section \ref{sec:observations} presents the \TESS{} photometry and the available optical, photometric, and spectroscopic observations of HD~191939. In Section \ref{sec:star_characterization}, we constrain the stellar parameters of HD~191939. Section  \ref{sec:ruling_out_false_positives} examines multiple false positive scenarios and confirms the planetary nature of the \TESS{} transit signals. In Section \ref{sec:analysis}, we describe our transit fitting routine, determine the system's physical and orbital parameters, investigate its dynamical properties, and discuss its prospects for atmospheric characterization. Section \ref{sec:discussion} places HD~191939 in the context of known planetary systems and highlights possible research avenues to improve our current knowledge of HD~191939. Finally, we summarize our results and present our conclusions in Section \ref{sec:conclusion}.

\pagebreak
\section{Observations} \label{sec:observations}
\subsection{\TESS{} Photometry}\label{sec:tess_lc}

With a \TESS{} magnitude of $T=8.29$\,mag, a radius smaller than the Sun's ($R_{*}=0.945\pm0.021\,R_{\oplus}$), and a low contaminating ratio ($\sim0.005$), HD~191939 was included in the \TESS{} Candidate Target List as a high-priority target \citep{TICv8}. As such, HD~191939 was pre-selected for 2-min. observations consisting of $11\times11$ pixels sub-arrays centered on the target. The star's astrometric and photometric properties are listed in Table \ref{tab:TOI_1339_stellar_params}. 

The \TESS{} spacecraft observed HD~191939 (RA J2015.5 = 20:08:06.150, Dec J2015.5 = +66:51:01.08) during Sectors 14--19 (2019 July 18 to 2019 December 24). After visually inspecting the target pixel files, we found that the host star had fallen outside of the CCD's science image area in Sector 14. We thus performed our analysis with data from sectors 15--19. 

\begin{figure*}[!htbp]
    \centering
    \hspace{-3em} 
    \begin{minipage}{\textwidth}
        \centering
        \includegraphics[width=0.85\textwidth]{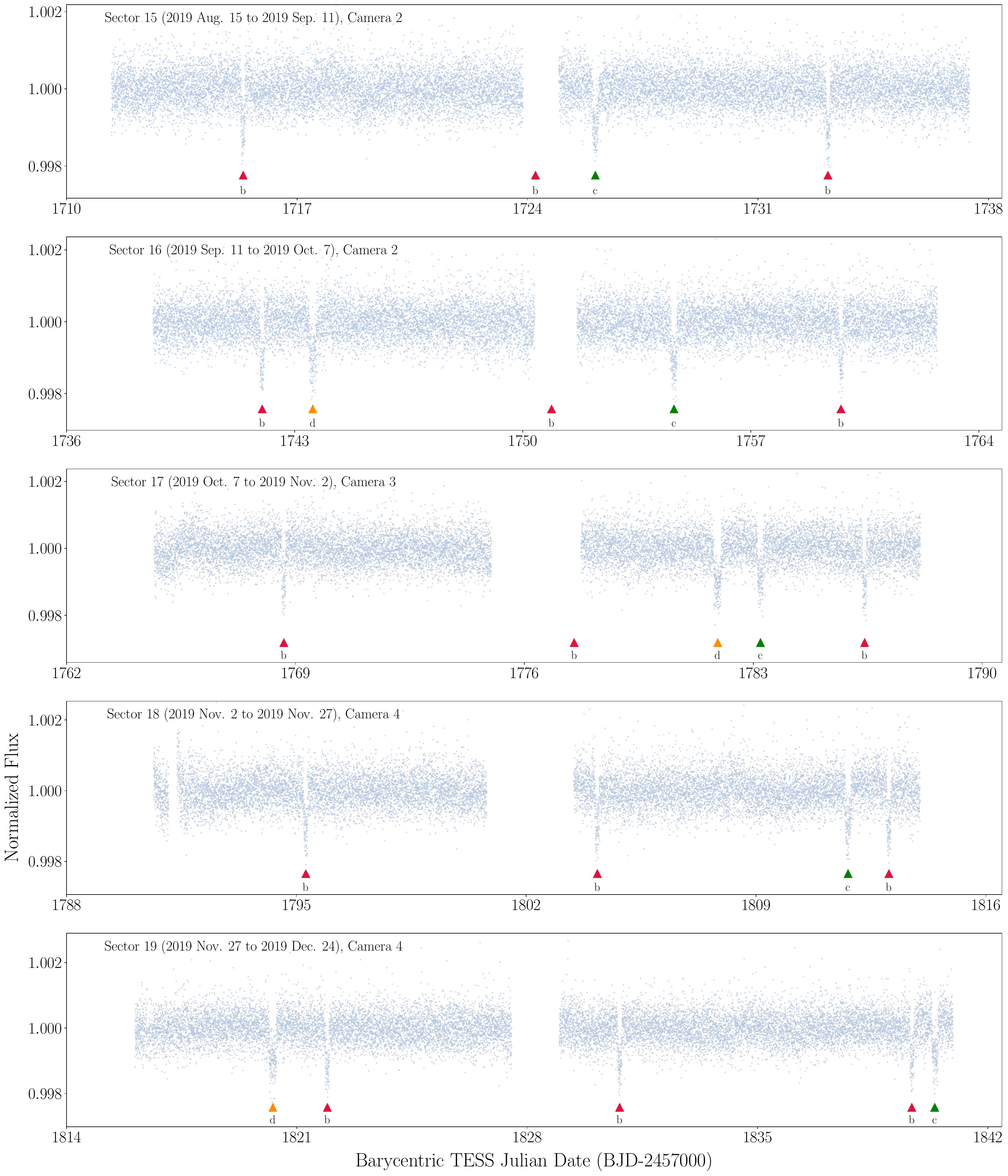}
    \end{minipage}
    ~
    \begin{minipage}{0.25\textwidth}
        \centering
        \includegraphics[width=\textwidth]{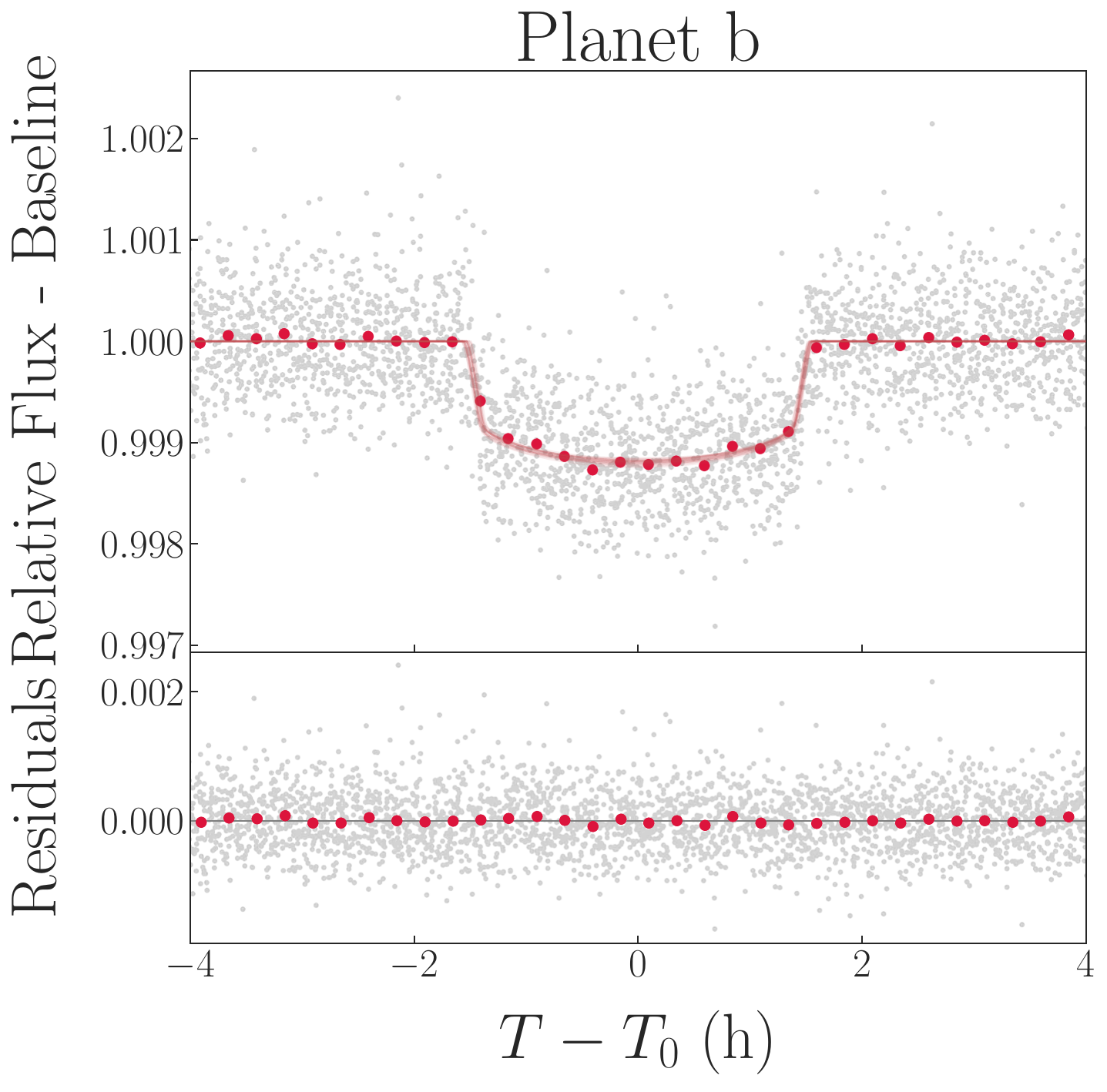}
    \end{minipage}
    ~ 
    \begin{minipage}{0.25\textwidth}
        \centering
        \includegraphics[width=\textwidth]{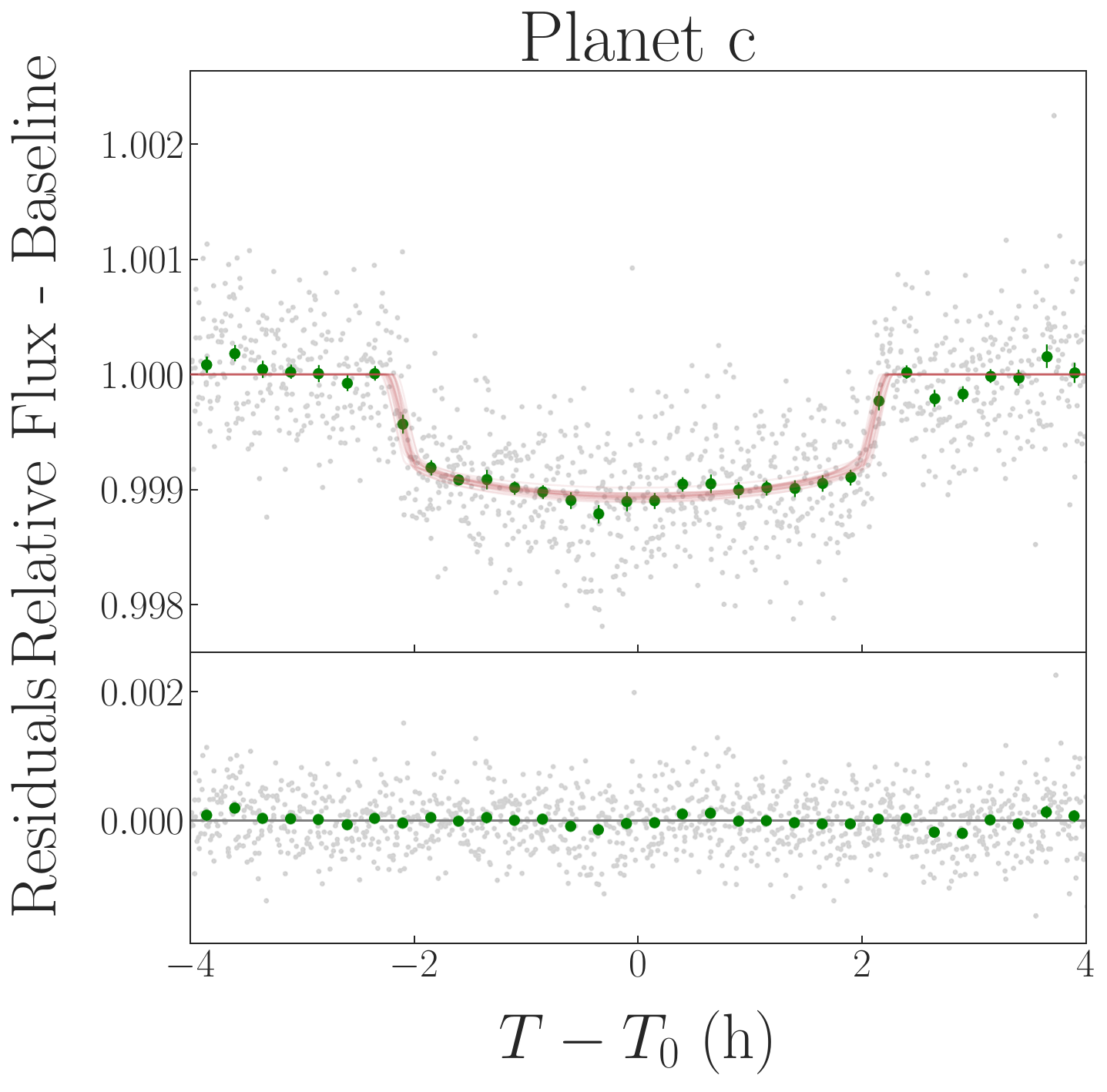}
    \end{minipage}
    ~
    \begin{minipage}{0.25\textwidth}
        \centering
        \includegraphics[width=\textwidth]{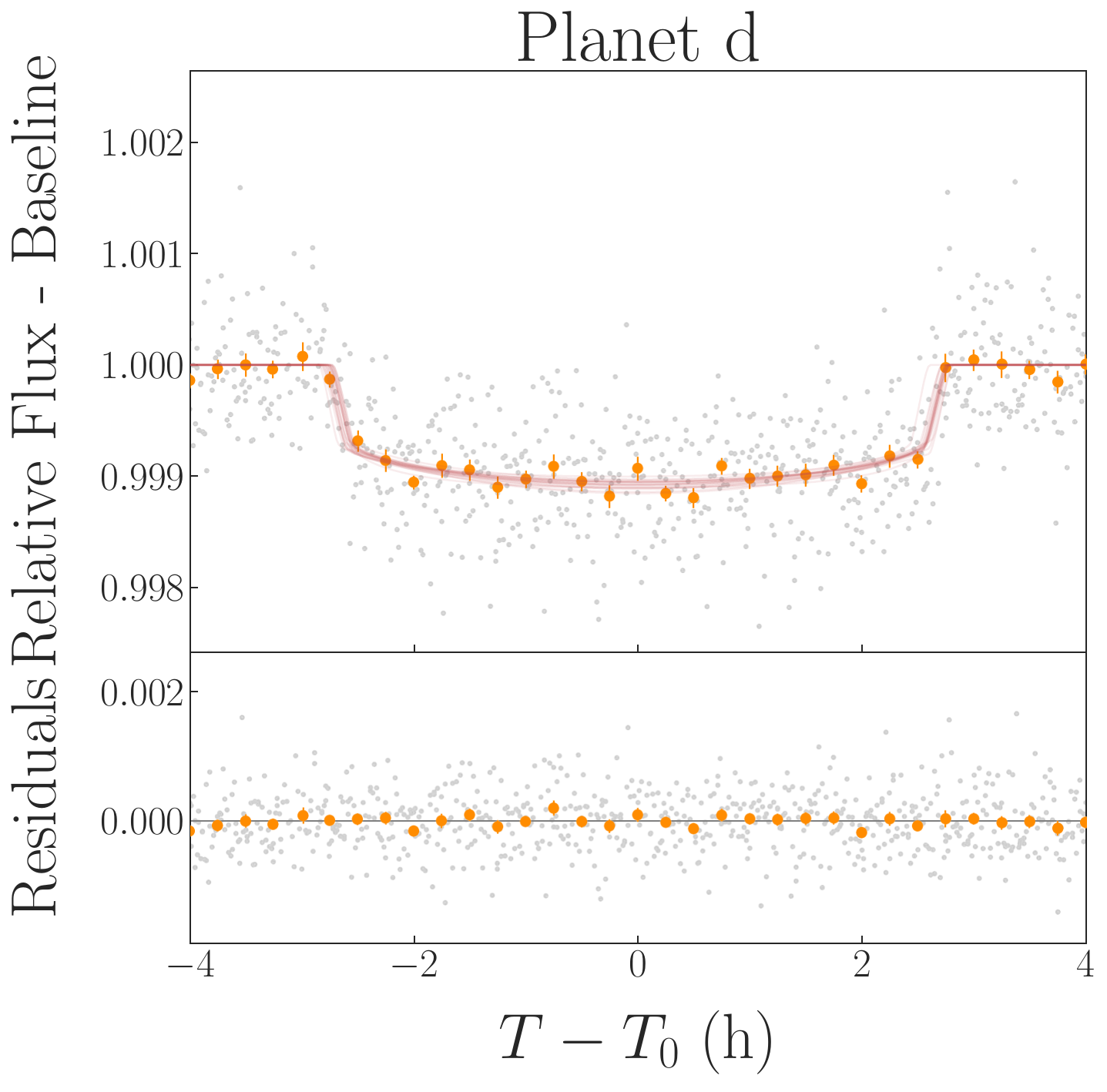}
    \end{minipage}
    \caption{\label{fig:lightcurves} \textit{Top}: The full \TESS{} discovery light curve  based on the 2-min. exposures from Sectors 15--19. The transits of planet b, c and d are shown in red, green, and orange, respectively. \textit{Bottom}: \TESS{} phase-folded light curves over the \allesfitter{} best-fit periods and initial epochs (see Section \ref{sec:analysis}). The grey points are the \TESS{} 2-min. exposures, the colored circles are the data points binned over 15 minutes, and the red lines represent 20 posterior models drawn from the outcome of the final fit. The light curve residuals are shown in the bottom panel.}
\end{figure*}

The photometric observations for HD~191939 (see \autoref{fig:lightcurves}) were  processed through the Science Processing Operations Center (SPOC) pipeline, developed and maintained by the NASA Ames Research Center \citep{Jenkins:2016, Jenkins2017}.\footnote{The SPOC pipeline searches for planetary transits by fitting an averaged Mandel $\&$ Agol \citep{MandelAgol:2002} model to the light curve with non-linear limb-darkening coefficients as parametrized by \citealt{ClaretBloemen:2011} \citep{Seader:2013, Li:2019}.} The pipeline detected two tentative planetary signals in the combined transit search for Sectors 15 and 16. With the addition of Sectors 17--19, the MIT Quick Look Pipeline (QLP; Huang et al., in prep.) identified three recurring transit signals. 

The phase-folded light curves obtained with the SPOC transit parameters had a flat-bottomed shape, consistent with a planetary interpretation of the transits. Moreover, the two planet candidates passed all the SPOC and QLP standard validation diagnostics, including a search for secondary eclipses, differences in odd and even transits, and flux centroid offsets during transit (see Section \ref{sec:SPOC_validation}). 

\begin{deluxetable}{lcc}[htbp]
    \tabletypesize{\small}
    \tablecaption{Stellar Properties of HD~191939. \label{tab:TOI_1339_stellar_params}}
    \startdata
     \\
    Property & Value & Source  \\
	\hline
	\multicolumn{3}{c}{\it Other Target Names}  \\
	\hline
	HD ID       &   191939               & 4    \\
	TOI ID      &   1339                 & -    \\
	TIC ID      &   269701147            & 1    \\
	2MASS ID    &   J20080574+6651019    & 2    \\
	Gaia DR2 ID &   2248126315275354496  & 3    \\
	\hline
	\multicolumn{3}{c}{\it Astrometric Properties}\\
	\hline 
    R.A. (J2015.5; h:m:s)	    	           &  20:08:06.150        & 3\\
    Dec	(J2015.5; d:m:s)	                   & +66:51:01.08         & 3\\
    Parallax (mas)           & $18.706\pm0.071$                 & 7\\
    $\mu_{{\rm R.A.}}$ (\masy) & $150.256\pm0.044$                    & 3 \\
	$\mu_{{\rm Dec.}}$ (\masy) & $-63.909\pm0.047$                    & 3 \\
    \hline
    \multicolumn{3}{c}{\it Photometric Properties}\\
	\hline
	\TESS{} (mag)       & $8.292\pm0.006$    & 1             \\
    B (mag)             & $9.720\pm0.038$    & 5             \\
	V (mag)             & $8.97\pm0.03$      & 6             \\
	Gaia (mag)          & $8.7748\pm0.0002$  & 3             \\
    J (mag)             & $7.597\pm0.029$    & 2             \\
    H (mag)             & $7.215\pm0.023$    & 2             \\
    K$_{s}$ (mag)       & $7.180\pm0.021$    & 2             \\
    \hline
    \enddata
   	\tablerefs{(1) \TESS{} Input Catalog Version 8 (TICv8) \citep{TICv8}. (2) Two Micron All Sky Survey (\textit{2MASS}; \citealt{2MASS_Survey}). (3) \Gaia{} DR2 \citep{GaiaDR2}. (4) Henry Draper Catalog \citep{HDCatalog}. (5) Tycho2 Catalog \citep{Tycho2Catalog}. (6) Hipparcos Catalog \citep{HipparcosCatalog}. (7) \Gaia{} DR2 parallax and uncertainty from TICv8, corrected for a systematic offset of $+0.082\pm0.033$\,mas, as described in \citealt{StassunTorres:2018}.}
\end{deluxetable}

We retrieved the SPOC-processed data from the \textit{Mikulski Archive for Space Telescopes} (MAST).\footnote{\url{https://mast.stsci.edu/portal/Mashup/Clients/Mast/Portal.html}} In particular, we downloaded the Presearch Data Conditioning (PDC) light curves, and removed all the observations encoded as \textit{NaN} or flagged as bad-quality points by the SPOC pipeline. From a total of 85282 photometric measurements (17848, 16812, 16945,  16612, and 17065 for Sectors 15--19, respectively), we identified a total of 4980 bad-quality data points, which we excluded from further analysis.

\subsection{Ground-Based Photometry: Observatori Astron\`omic Albany\`a}\label{sec:phot_follow_up}

As part of the TFOP follow-up program,\footnote{\url{https://tess.mit.edu/followup/}} we acquired 320 photometric exposures of HD~191939 on 2019 October 29 with the 0.4-m telescope at the Observatori Astronòmic Albanyà (OAA) in Catalonia (Spain). The host star was continuously observed for 398.8 minutes in the Cousins $I_{c}$ filter using a CCD camera with a resolution of 3056$\times$3056 pixels and a pixel scale of $0.72\arcsec$ per pixel. The science exposures were reduced with the AstroImageJ (AIJ) software \citep{Collins+2016}.

\subsection{Archival Spectroscopic Observations: SOPHIE} \label{sec:SOPHIE} 

\SOPHIE{} \citep{Bouchy:2009, Perruchot:2008} is a fiber-fed échelle spectrograph mounted on the 1.93-m telescope at the Observatoire de Haute Provence (OHP), in France. This instrument observed HD~191939 between 2007 September 27 and 2007 November 30 with a RV precision of 4--5~$\mathrm{m\,s^{-1}}$ \citep[e.g.][]{2009A&A...505..853B,2011epsc.conf..240B}. A total of five spectra were acquired with \SOPHIE{}'s high-resolution mode, which provides a resolving power of $\lambda/\Delta\lambda\equiv R = 75,000$. The spectra have a median exposure time of $617$\,seconds and a median SNR per pixel at 550\,nm of 59 (see Table \ref{tab:SOPHIE_data}).

\begin{table}[htbp]
    \centering
    \caption{\SOPHIE{} RV measurements with their SNR at 555\,nm and their exposure times.}
    \begin{tabular}{ccccc}
        \hline
        \hline
        \noalign{\smallskip}
        BJD$_\mathrm{UTC}$ & RV  & Error & SNR & Exp. time  \\
         $-2,450,000$ & (\kms) & (\kms) & - & (s)\\ 
        \hline
        4371.345 & -9.237 & 0.002 & 42.9 & 600 \\ 
        4372.288 & -9.249 & 0.001 & 60.5 & 500 \\
        4430.266 & -9.213 & 0.001 & 70.8 & 900 \\
        4431.296 & -9.218 & 0.001 & 67.4 & 743 \\
        4435.308 & -9.232 & 0.001 & 54.4 & 346 \\
        \hline
        \multicolumn{4}{l}{\begin{minipage}{3.1in}
        \end{minipage}}\\
    \end{tabular}
    \label{tab:SOPHIE_data}
\end{table}

We downloaded all the available observations of HD~191939 from the \SOPHIE{} public archive \citep{2004sf2a.conf..547M}. These included spectra reduced by the \textit{Data Reduction Software v0.50}\footnote{\url{http://www.obs-hp.fr/guide/sophie/data_products.shtml}.} (DRS), as well as the cross-correlation functions (CCFs) determined by the DRS using a numerical mask for the G2 spectral type \citep{1996A&AS..119..373B}. The CCFs were calculated over a $\pm30$\,km/s velocity interval. The RV, full-width half maximum (FWHM), and contrast of each CCF were computed by the DRS by fitting a Gaussian function to the CCF profile. We extracted barycentric-corrected radial velocities, FWHM, and the bisector spans from the FITS headers of the CCFs \citep{SOPHIE_DRS}. We found no correlations between either the RV measurements and the bisectors ($r$ = 0.51, while the critical value of the Pearson correlation coefficient at the confidence level of 0.01 is $r_{3,0.01}$ = 0.96), or between the RVs and the FWHM of the CCFs ($r$ = 0.43). Such correlations would have indicated astrophysical false positives, such as stellar spots or blends. 

\subsection{Spectroscopic Follow-up} \label{sec:spec_follow_up}
\subsubsection{TRES Reconnaissance Spectroscopy} \label{sec:TRES}

We obtained three spectra of HD~191939 between 2019 October 24 and 2019 November 5 using the fiber-fed Tillinghast Reflector Echelle Spectrograph (TRES; \citealt{Furesz:2008}) on the 1.5-m telescope at the Fred Lawrence Whipple Observatory (Mt. Hopkins, Arizona). TRES covers the spectral range 3850--9100\,$\AA$ and has a resolving power of $R=44,000$. The TRES spectra have an average SNR per resolution element of 47 and were extracted as described in \citet{Buchhave:2010}.

\subsubsection{LCO/NRES Reconnaissance Spectroscopy} \label{sec:NRES}

We acquired three consecutive 20-minute optical exposures of HD~191939 on 2019 November 1 with the Network of Echelle Spectrographs (NRES; \citealt{Siverd:2016, Siverd:2018}), operated by Las Cumbres Observatory (LCO; \citealt{Brown:2013}). The NRES is composed of four high-precision fiber-fed spectrographs with a resolving power of $R=53,000$. The NRES spectra were stacked to remove cosmic rays and instrumental effects, resulting in a final SNR of 60. 

\subsection{Gemini/NIRI High Angular Resolution Imaging}\label{sec:gemini_niri}

\begin{figure}[!htbp]
    \centering
    \includegraphics[width=0.98\columnwidth]{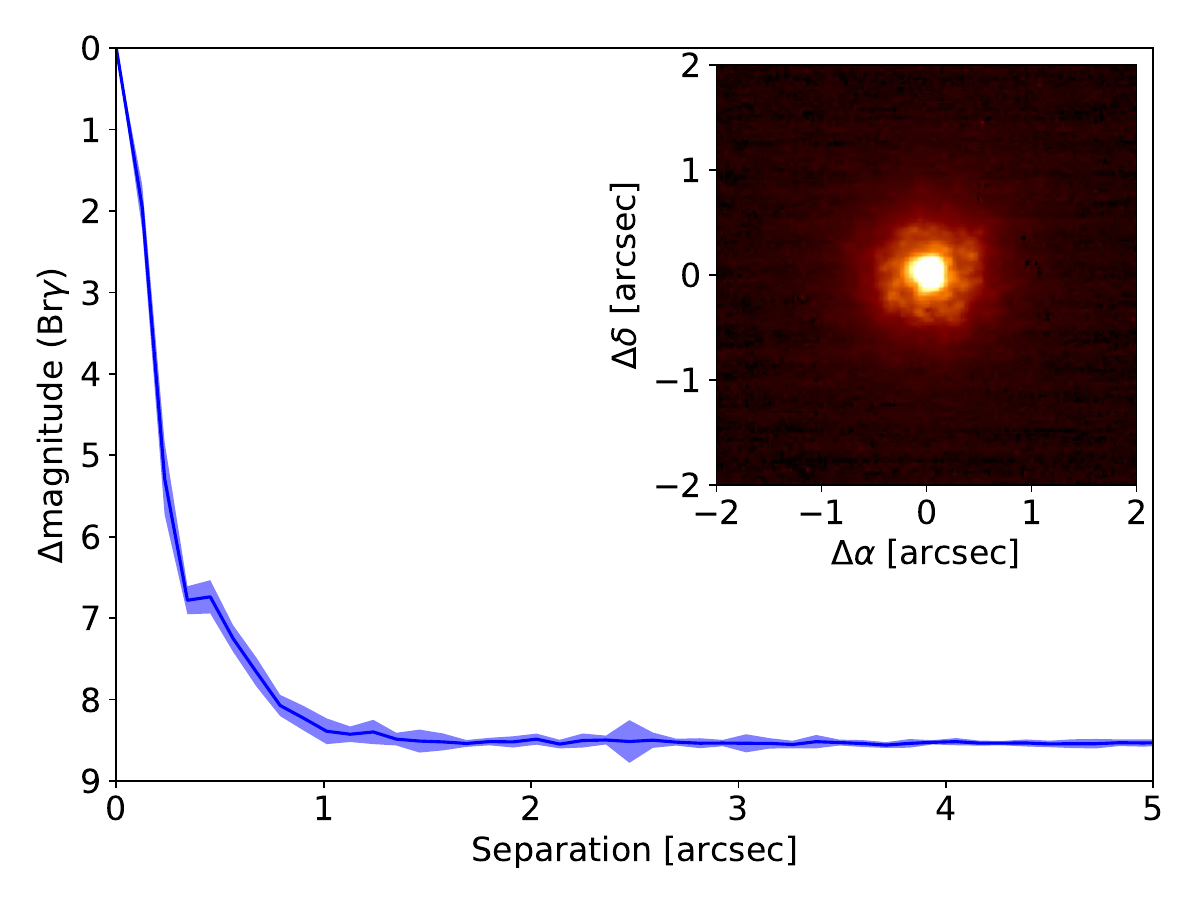}
    \caption{Sensitivity curve of our Gemini/NIRI AO images (solid black line). We are sensitive to companions with a contrast of 5\,mag just $200$\,mas from the star. No visual sources are seen anywhere in the field-of-view. A thumbnail image of the target is inset.}
    \label{fig:AO_image}
\end{figure}

We observed HD~191939 on 2019 November 8 with the Near InfraRed Imager (NIRI) at the Gemini North telescope \citep{GeminiNIRI:2003}. A total of nine Adaptive Optics (AO) images were collected in the Br$\gamma$ filter, each with an exposure time of 2 seconds. The telescope was dithered in a grid pattern between each science exposure to construct the sky background frame and remove artifacts such as bad pixels and cosmic rays. Data were processed using a custom set of IDL codes with which we interpolated bad pixels, subtracted the sky background, flat corrected images, aligned the stellar position between frames, and co-added data. We determined the sensitivity to stellar companions by injecting artificial point spread functions into the data at a range of separation and angles, and scaling these until they could be detected at 5$\sigma$. We are sensitive to stellar neighbors 5 magnitudes (8.4 magnitudes) fainter than HD~191939 at $200$\,mas ($1\arcsec$). Our sensitivity as a function of radius is shown in \autoref{fig:AO_image} with a thumbnail image of HD~191939.

\section{Host Star Characterization}\label{sec:star_characterization} 

\begin{figure*}[!htbp]
    \centering
    ~
    \begin{minipage}{0.465\textwidth}
    \centering
    \includegraphics[width=\textwidth]{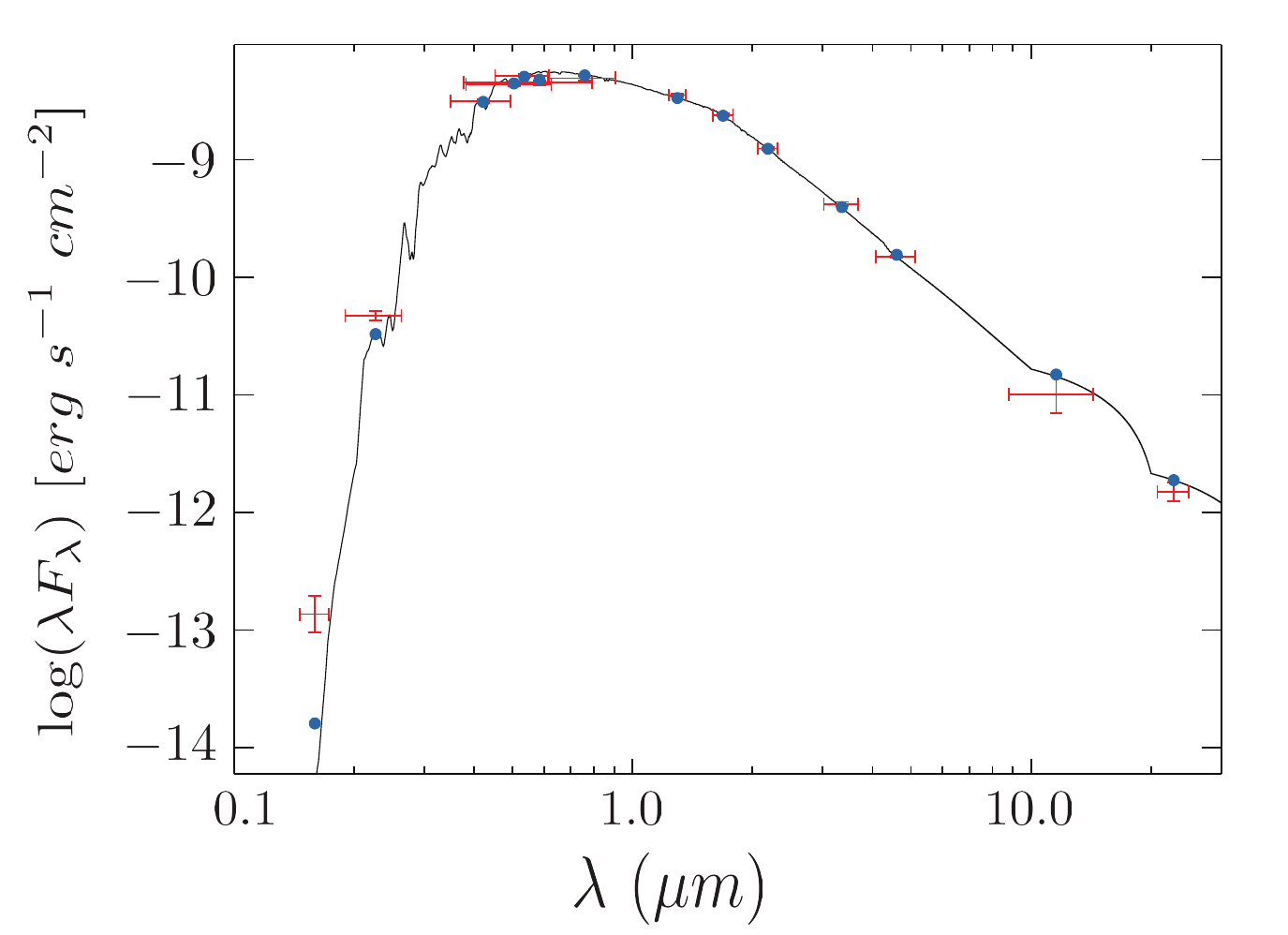}
    \end{minipage}    
    ~
    \begin{minipage}{0.435\textwidth}
        \centering
        \includegraphics[width=\textwidth]{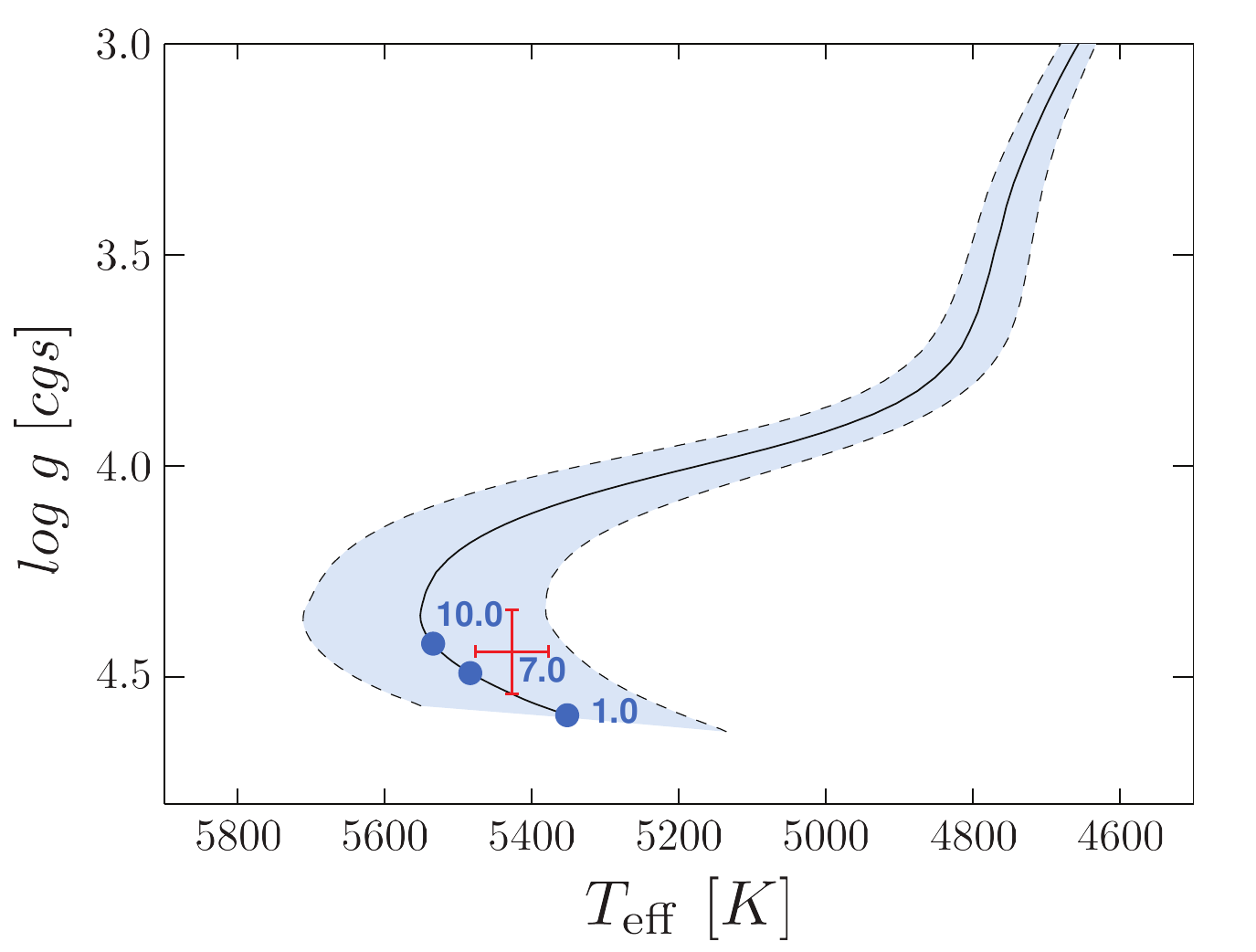}
    \end{minipage}
    \caption{
    \label{fig:sed_hdr}
    \textit{Left}: Spectral energy distribution (SED). Red symbols are the observed photometric data, with the horizontal bars reflecting the effective width of the passband. Blue symbols are the model fluxes from the best-fit Kurucz atmosphere model (black). \textit{Right}: H-R diagram. The black curve with blue swathe represents a Yonsei-Yale evolutionary model for the stellar mass and metallicity with their uncertainties. The blue dots label ages along the evolutionary track in Gyr. The red symbol represents the spectroscopically derived effective temperature and surface gravity with their uncertainties.}
\end{figure*}

\subsection{TRES Spectroscopy} \label{sec:TRES_stellar_characterization}

We used the TRES spectra to measure the host star's effective temperature $T_\mathrm{eff}$, surface gravity $\log{g}$, metallicity $\left[\text{m/H}\right]$, and rotational velocity $v \sin{i}$ with the Stellar Parameter Classification (SPC; \citealt{Buchhave:2012}) tool. The SPC software cross-correlates an observed spectrum against a grid of synthetic spectra based on the Kurucz atmospheric models \citep{Kurucz:1992}. The weighted average results are $T_{\rm eff} = 5427\pm50$\,K, $\log g = 4.44 \pm0.10$\,cgs, [m/H] = $-0.16 \pm0.08$\,dex, and $v\sin{i}= 0.6 \pm0.5$\,km/s (see Table~\ref{tab:derived_TOI_1339_stellar_params}). From the TRES spectra, we also detected weak H-alpha absorption indicating that HD~191939 has a low stellar activity. 

\subsection{NRES Spectroscopy} \label{sec:NRES_stellar_characterization}

We also constrained the stellar parameters from the NRES observations following the techniques presented in \citet{Petigura:2017} and \citet{Fulton:2018}. In particular, we used the \texttt{SpecMatch} software \citep{Petigura:2015_phd,  Petigura:2017}\footnote{\url{https://github.com/petigura/specmatch-syn}} to compare the observed spectrum of HD~191939 against synthetic spectra created by linearly interpolating the \citet{Coelho:2005} grid of model spectra at arbitrary sets of \teff, \logg, $\left[\text{Fe/H}\right]$, and $v\sin{i}$. We maximized the $\chi^2$-based likelihood via a Levenberg-Marquardt algorithm (\citealt{Press:1992}) and found \teff = $5335\pm100$\,K, \logg = $4.2\pm0.1$\,cgs, $\left[\text{Fe/H}\right]$ = $-0.13\pm0.06$\,dex, and $v \sin{i} < 2$\,km/s. 

\subsection{Spectral Energy Distribution Analysis} \label{sec:SED}

We used the host star's broadband Spectral Energy Distribution (SED) and its \Gaia{} DR2 parallax to determine an empirical measurement of the stellar radius following procedures described in the literature \citep{Stassun:2016, Stassun:2017, Stassun:2018}. For this analysis, we retrieved the FUV and NUV fluxes from \textit{GALEX}, the $B_T$ and $V_T$ magnitudes from \textit{Tycho-2}, the $J$, $H$, and $K_{s}$ magnitudes from \textit{2MASS}, the W1--W4 magnitudes from \textit{WISE}, and the $G$, $G_{\rm BP}$, and $G_{\rm RP}$ magnitudes from \textit{Gaia}. When taken in combination, the available photometry spans the full stellar SED over the wavelength range 0.15--22\,$\micron$ (see \autoref{fig:sed_hdr}). 

We performed a fit to the host star's SED with the Kurucz stellar atmospheric models, placing priors on \teff, \logg, and  $\left[\text{m/H}\right]$ based on the SPC analysis of the TRES spectra. The remaining free parameter was the extinction ($A_V$), which we limited to the maximum line-of-sight value from the dust maps of \citet{Schlegel:1998}. The model fits the data well, with a reduced $\chi^2$ of 1.9 and best-fit extinction of $A_V = 0.03 \pm 0.03$ (see \autoref{fig:sed_hdr}). We also integrated the model SED to obtain a bolometric flux at Earth of $F_{\rm bol} = (7.81 \pm 0.18) \times 10^{-9}$\, erg~s$^{-1}$~cm$^{-2}$. Using $F_{\rm bol}$, $T_{\rm eff}$, and the \Gaia{} DR2 parallax adjusted by $+0.08$\,mas to account for the systematic offset reported by \citet{StassunTorres:2018}, we determined a stellar radius of $R_{*} = 0.945 \pm 0.021$\,R$_\odot$. We also estimated the stellar mass empirically. The eclipsing binary-based relations of \citet{Torres:2010} yield $M_\star = 0.92 \pm 0.06$\,M$_{\odot}$, whereas the stellar surface gravity and SED-based radius result in $M_\star = 0.90 \pm 0.21$\,M$_\odot$. \autoref{fig:sed_hdr} shows the former in a Hertzsprung-Russell (H-R) diagram with an evolutionary track from the Yonsei-Yale models \citep{Yi:2001,Spada:2013}. These plots imply that the age of HD~191939 is $7\pm3$\,Gyr. 

Finally, we used the spectroscopic $v\sin{i}$ and the SED-based radius to calculate a stellar rotation period of $P_{\rm rot}/\sin i = 79 \pm 66$\,d, where the large uncertainty is driven by the large error on the spectroscopic $v\sin{i}$. 
This is consistent with the dominant periodicity in the \TESS{} data (after masking the transits of the three planets) identified via a Lomb-Scargle periodogram analysis \citep{Lomb1976, Scargle1982}: a 44-ppm peak-to-peak roughly sinusoidal variation with a 14.15-day period and false alarm probability of 10$^{-20}$.
\subsection{Independent Validation of Stellar Parameters}  \label{sec:iso_packages}  
\begin{table*}[!ht]
    \centering
    \caption{Derived stellar properties for HD~191939.
    }
    \begin{tabular}{lccc}
    \hline
    \hline
	Property & Value & Source  & Reference spectra \\
	\hline
	$R_{*}$ (\Rsun) 		          &  $0.945\pm0.021$                   &  SED                                  & TRES \\
    $M_{*}$ (\Msun) 		          &  $0.92\pm0.06$                     &  SED via \citealt{Torres:2010}        & TRES \\
    Age (Gyr)                         &  $7\pm3$                           & SED                                                                                         & TRES \\ 
    $A_{v}$                           &  $0.03\pm{0.03}$            &  SED                                                                                        & TRES \\
    $F_{\text{bol}}$ (erg/s/cm$^{2}$) &  $(7.81 \pm 0.18) \times 10^{-9}$  &  SED                                                                                        & TRES \\
    $T_\mathrm{eff}$ (K)              &  $5427\pm50$                       &  SPC                                                                                        & TRES \\
    $\log{g}$ (cgs)                   &  $4.40\pm0.10$                     &  SPC                                                                                        & TRES \\
    $\left[m/H\right]$ (dex)	      &	 $-0.16\pm0.08$                    &  SPC                                                                                        & TRES \\
    $(v \sin{i})_\mathrm{A}$ (\kms)	  &  $0.6\pm0.5$                       &  SPC                                                                                        & TRES  \\
    $L_{*}$ (L$_{\odot}$)             &  $0.69\pm{0.01}$            & \isochrones{} (\textbf{M}ESA \textbf{I}sochrones $\&$ \textbf{S}tellar \textbf{T}racks)     & TRES  \\ 
    Distance (pc)                     &  $53.48^{+0.19}_{-0.20}$           & \isochrones{} (\textbf{M}ESA \textbf{I}sochrones $\&$ \textbf{S}tellar \textbf{T}racks)     & TRES \\ 
    $\rho$ (\gccc) 		              &  $1.55\pm{0.19}$            &  \allesfitter{}                       &  \\
    Spectral type                     &           G8 V                     &  \cite{Pecaut2012}, \cite{Pecaut2013} &  \\
    $W_{H\alpha}$ ($\mathrm{\AA}$)    & $1.259\pm0.007$                    & This Work                             & \SOPHIE{} \\
	\hline
    \multicolumn{4}{l}{\begin{minipage}{3.1in}
    \end{minipage}}\\
    \end{tabular}
    \label{tab:derived_TOI_1339_stellar_params}
\end{table*}

As an independent validation on the SPC/SED stellar parameters, we used the spectroscopic properties of HD~191939  derived from the TRES and NRES spectra to perform isochrone fitting with two stellar evolutionary models: the MESA Isochrones and Stellar Tracks database (MIST; \citealt{Choi:2016, Dotter:2016}) as implemented by the \isochrones{} \citep{Morton:2015} and \isoclassify{} \citep{Huber:2017} packages, and the Padova models (\citealt{Padova_DaSilva:2006}), accessible via the \texttt{PARSEC v1.3} \citep{PARSEC:2012} online tool.\footnote{\url{http://stev.oapd.inaf.it/cgi-bin/param_1.3}} 

We ran \isochrones{} and \isoclassify{} with priors on the star's photometric magnitudes, the corrected \Gaia{} DR2 parallax, and the best-fit spectroscopic parameters from either the TRES (\teff, \logg, $\left[\text{m/H}\right]$) or NRES (\teff, \logg, $\left[\text{Fe/H}\right]$) spectra.\footnote{For the TRES spectra, we assumed that  $\left[\text{m/H}\right]$ was a good first-order initial guess for $\left[\text{Fe/H}\right]$, as Sun-like stars such as HD~191939 are not particularly enriched in alpha elements.} To implement \isoclassify{}, we also accounted for extinction by incorporating the 3D dust map of \cite{Green:2018}, which covers most of the sky with a declination larger than $-30$\textdegree. The derived stellar parameters are consistent with the values presented in Table \ref{tab:derived_TOI_1339_stellar_params}, regardless of the choice of reference spectra for the host star's spectroscopic parameters. To assess whether this consistency was primarily due to the use of the MIST database, we also determined the stellar parameters with the Padova models with the \texttt{PARAM} v1.3 tool. For both the TRES and NRES spectroscopic parameters, the resulting stellar properties agree with the \isochrones{} and \isoclassify{} predictions to within 1$\sigma$. In the rest of this paper, we adopt the results from Table \ref{tab:derived_TOI_1339_stellar_params} for our analysis of the HD~191939 system.

\pagebreak
\subsection{Chromospheric Activity Indicators} \label{sec:chromospheric_analysis}

The \SOPHIE{} spectra detailed in Section~\ref{sec:SOPHIE} indicate that HD~191939 is chromospherically inactive. As with the TRES spectra, the five \SOPHIE{} spectra show the H$\alpha$ line in absorption. We measured the equivalent width of the H$\alpha$ line ($W_{H\alpha}$) in each spectrum using a 10-$\mathrm{\AA}$ subsample centered on the vacuum wavelength of H$\alpha$ (6562.81\,$\mathrm{\AA}$). We fitted a Voigt profile to the line and a linear trend to the continuum via least squares using \texttt{astropy} \citep{astropy:2018}. We estimated the uncertainties by bootstrapping the model fit 100 times excluding a random 10\% of the data points. 
The equivalent widths are consistent between the five spectra, and we measure a weighted-mean H$\alpha$ equivalent width of $W_{H\alpha} = 1.259\pm0.007\,\mathrm{\AA}$. Visual inspection reveals no evidence of emission in the cores of the Ca~II H and K or H$\alpha$ lines. We conclude these factors indicate a lack of measurable chromospheric activity for HD~191939.

\section{Ruling out False Positives}\label{sec:ruling_out_false_positives} 

\begin{figure*}[!htbp]
\centering
\includegraphics[width=\textwidth]{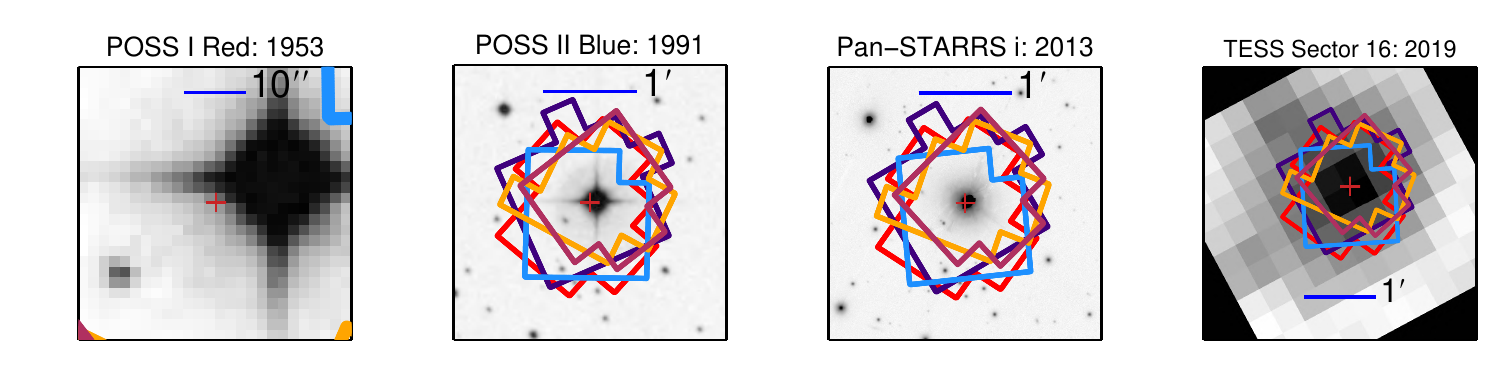}
\caption{\label{fig:archival_images}
Archival images and \TESS{} view of the field around HD~191939 from 1953 to 2019. North points up and East is to the left. The red cross is the star's current location, and the red, purple, light blue,  orange, and maroon regions are the \TESS{} photometric apertures for Sectors 15, 16, 17, 18, and 19 respectively. Due to the star's proper motion, there is an offset between its present-day position and its location in the original POSS images.}
\end{figure*}

The \Kepler{} mission revealed that multiple-period transit-like events are more likely to be caused by true planets than by false positives (e.g. \citealt{Latham:2011, Lissauer2012}). Despite such evidence, it is  important to carefully inspect the observed transit signals to rule out false positive scenarios, including instrument systematics and contamination from nearby stars. In this section, we aim to validate the HD~191939 planet candidates with the SPOC Validation Tests (Section \ref{sec:SPOC_validation}), the \TESS{} photometry and archival/follow-up observations of HD~191939 (Section \ref{sec:TFOP_validation}), and the statistical validation software \VESPA{} (Section \ref{sec:VESPA_validation}). 

\subsection{SPOC Validation Diagnostics}\label{sec:SPOC_validation}

The two planet candidates identified by the SPOC Data Validation Pipeline (referred to as planet "b" and "c" in our analysis; see Section \ref{sec:analysis}) pass all the SPOC Data Validation (DV) tests \citep{Twicken:2018}. We list these tests in the following paragraphs. 

\begin{itemize}[noitemsep,topsep=0pt,leftmargin=12pt]
    \item An \textit{Eclipsing Binary Discrimination Test} to search for weak secondary eclipses and compare the depth of odd and even transits. Planet b and c pass this diagnostic at 2$\sigma$, with no shallow secondaries around phase 0.5 and no odd/even transit depth variations.
    \item An Optical \textit{Ghost Diagnostic Test} designed to primarily rule out optical ghosts, scattered light, instrumental noise, and bright background EBs (outside of the photometric aperture) as the source of the transit-like events. This test measures the correlation between a transit model light curve and flux time series derived from the photometric core and halo aperture pixels to determine whether the transit signature is more consistent with (1) a star in the photometric core,  or (2) distributed or other contamination outside the core. Planets b, c and d all pass this diagnostic test within 2$\sigma$.
    \item A \textit{Difference Image Centroid Offset Test} to determine if the location of the transit source is statistically consistent with the position of the target star. The offset distance for planet b in the combined Sector 15--16 SPOC analysis was less than $1.7\arcsec$ (0.33$\sigma$). For planet c, the maximum offset distance was less than $5.2\arcsec$ (0.94$\sigma$). 
    \item A \textit{Bootstrap Test} to assess the confidence level of the transit detection. Planet b and c pass this test with formal false alarm probabilities of $1.05\times10^{-135}$ and $1.72\times10^{-63}$, respectively.
\end{itemize}

These DV tests were not applied to the third planet around HD~191939 (or planet "d"), as it was not detected by the SPOC pipeline in the combined transit search of \TESS{} Sectors 15 and 16. Nevertheless, we independently verified the planetary nature of all three planet candidates with the analyses described in the following sections.

\subsection{Observational Constraints} \label{sec:TFOP_validation}

\subsubsection{Archival Optical Images}  \label{sec:archival_optical_images}

The \TESS{} detectors have a larger pixel scale than the \Kepler{} telescope (\TESS{}: $\sim21''$, \Kepler{}: $4''$), so photometric contamination from nearby astrophysical sources is more likely. To investigate this false positive scenario, we compared a \TESS{} exposure of HD~191939 from Sector 16 to archival optical images taken in 1953, 1991, and 2013 by the first Palomar Observatory Sky Survey (POSS-I; \citealt{POSS-1:1963}), the second POSS (POSS-II; \citealt{POSS-2:1991}) and the Pan-STARRS survey \citep{PanSTARRS:2002, PanSTARRS:2010}, respectively (see \autoref{fig:archival_images}). Due to HD~191939's high proper motion, its present-day location appears unobscured in the archival images. Based on the POSS-I field-of-view, when HD~191939 was $\sim11''$ away from its current sky position due to proper motion, we estimate that any $V\lesssim19$ stars would have been clearly visible where HD~191939 is located today (e.g. a $V=18.34$ source, identified as \Gaia{} DR2 2248126310978337408, can be observed in the bottom left corner of the POSS-I view). 

We performed a query of \Gaia{} DR2 and \textit{2MASS} catalogs centered on HD~191939, and used the SPOC reports to identify potential background sources around the host star. Within the central \TESS{} pixel, there is only one \textit{2MASS} source at 13.6$\arcsec$ separation with a \TESS{} magnitude of $T=14.70$\,mag (2MASS~J20080397+6651023; TIC~269701145). However, this object is likely a spurious \textit{2MASS} detection. First, \textit{2MASS} artifacts are known to appear around bright stars, typically along their diffraction spikes \citep{TICv8}. Second, it was only observed in the J-band ($J = 14.2$\,mag) and not in the H- and K-band (\textit{2MASS} photometric quality flag of "AUU" and read flag of 0 for H- and K-bands; \citealt{2MASS_Survey}). Third, it is likely that \Gaia{} DR2 would have detected this \textit{2MASS} object provided it were real. With a \Gaia{} magnitude difference of $\Delta G \sim5.4$\,mag relative to HD~191939,\footnote{The \textit{2MASS} detection lacks optical photometry, so there is no reliable way of estimating its \Gaia{} magnitude. To calculate $\Delta G$, we assume that this object is very red (e.g. a late M-dwarf) and take its \textit{2MASS} J-band magnitude as an approximation of its \Gaia{} magnitude. This conservative approach provides an estimate of how faint the \textit{2MASS} source could be in the \Gaia{} band-pass.} this source would have been within the observable parameter space of \Gaia{} DR2's contrast sensitivity curve \citep{Brandeker:2019}. Fourth, the Pan-STARRS images for HD~191939 in the \textit{grizy} filters (e.g. see \autoref{fig:archival_images}) do not reveal any stellar objects near the position of the \textit{2MASS} source . Given the high level of completeness of the Pan-STARRS survey down to its limiting magnitude (\textit{grizy}$\sim22.4$), it is thus improbable that the \textit{2MASS} source is a true star. Finally, we can rule out the existence of this artifact with ground-based photometry (see Section \ref{sec:OOA_EB_Analysis}). With all the aforementioned evidence, we conclude that the \textit{2MASS} source is an instrumental artifact and could not have caused the transit-like events in the \TESS{} light curve. 

\subsubsection{Ground-based Photometry}\label{sec:OOA_EB_Analysis}

The OAA observations covered a full transit of the inner planet HD~191939\,b, and showed a possible detection of a roughly 1200 ppm transit within a $13\arcsec$ photometric aperture. However, the data was not of sufficient quality to include in our global model fit. The longer periods of the planet candidates c and d have prevented successful ground-based photometric follow-up of their transits thus far.

To rule out nearby EBs, we conducted aperture photometry of all the visible sources within $2.5'$ of HD~191939 using a photometric aperture radius of $13\arcsec$. For each source, we employed the AIJ software to determine the root mean square error of its light curve, the predicted transit depth on the target star, and the resulting SNR. None of the sources considered in this analysis are bright enough to be a potential source of the \TESS{} detection. Moreover, the OAA exposures show no evidence of the apparent \textit{2MASS} instrumental artifact discussed in Section \ref{sec:archival_optical_images}.

\subsubsection{High-Resolution Imaging} \label{sec:gemini_niri_validation_test}

It is important to check for stellar companions that can dilute the light curve, thus biasing the measured planetary radius or even be the source of false positives (\citealt{Ciardi:2015}). To search for such companions, we examined the AO Gemini/NIRI final image (see Section \ref{sec:gemini_niri}) and found no visible stellar objects in the field-of-view around HD~191939 (\autoref{fig:AO_image}). 

\subsubsection{Archival Radial Velocities}  \label{sec:archival_rv_analysis}

We performed a joint fit to the \SOPHIE{} RV and \TESS{} observations with the \allesfitter{} package \citep{allesfitter}\footnote{\url{https://github.com/MNGuenther/allesfitter}} to rule out possible substellar or stellar companions. Our combined fit finds a $3\sigma$ upper limit on the RV semi-amplitudes ($K$) of $K_{b}<250$\,m/s,~$K_{c}<300$\,m/s, and $K_{d}<250$\,m/s for planet candidate b, c, and d, respectively. In contrast, a brown dwarf ($M\approx13\,M_{\text{Jup}}$) around HD~191939 would have $K_{b}=620$\,m/s,  $K_{c} = 420$\,m/s, and $K_{d}=380$\,m/s. The \SOPHIE{} constraints lie well below these values, thus pointing to the planetary origin of the \TESS{} transits. 

In addition, we estimated the RV semi-amplitudes of the HD~191939 planet candidates from the standard RV equation (see Eq. 14 in \citealt{Seager:2010_LovisRV}) using the \allesfitter{} orbital results (see Table \ref{tab:MCMC_params}) and the planets' masses predicted via the probabilistic mass-radius (MR) relation of \citealt{Wolfgang:2016} (W16). To estimate the HD~191939 planetary masses, we used the full \allesfitter{} posterior for the planetary radii and samples from the posterior of the parameters that define Eq. 2 in W16. This yields  $M_{b}=14.77^{+1.98}_{-1.97},M_{\oplus}$, $M_{c}=13.85
^{+1.87}_{-1.85}\,M_{\oplus}$ and $M_{d}=13.50^{+1.84}_{-1.80}\,M_{\oplus}$ for planet candidates b, c, and d, respectively. 
In turn, these masses correspond to RV semi-amplitudes ($K_{b}=2.0\pm0.6$ \,m/s, $K_{c} = 1.0\pm0.4$\,m/s, and $K_{d}=1.0\pm0.4$\,m/s), well within the range of the \SOPHIE{} predictions. We also estimated the planetary masses and RV semi-amplitudes with the probabilistic MR relation of \citealt{chen2017} and found consistent results.

\subsubsection{Ingress/Egress Test} 

We investigated whether a chance-aligned background or foreground EB could have caused the observed transits in the \TESS{} light curve by placing an  upper limit on the magnitude of a fully blended star. In a scenario of photometric contamination by blended light, the observed \TESS{} transit depth ($\delta_{\text{obs}}$) is given by:
\begin{equation}
    \centering
   \label{eq:delta_obs}
    \delta_{\text{obs }}\simeq\left(\frac{R_{p,\text{true}}}{R_{*}}\right)^{2}\frac{F_{\text{blend}}}{F_{\text{blend}}+F_{\text{star}}}=\delta_{\text{true}}\frac{f}{1+f},
\end{equation}
where $f$ is the flux ratio $f\equiv F_{\text{blend}}/F_{\text{star}}$, $\delta_{\text{true}}$ is the square of the true planet-to-star radius ratio in the absence of a blend, $F_\text{blend}$ is the flux of the contaminating source, and $F_{\text{star}}$ is the flux of HD~191939. Under the assumption of a central transit (i.e. $b=0$), the observed transit depth $\delta_{\text{obs}}$ must satisfy (Eq. 21 in \citealt{Seager:2003}):  
\begin{equation}
   \centering
   \label{eq:ratio_of_times}
   \delta_{\text{obs}}\leq\delta_{\text{blend}}=\
   \frac{\left(1-\frac{t_{F}}{t_{T}}\right)^{2}}{\left(1+\frac{t_{F}}{t_{T}}\right)^{2}},
 \end{equation}
where $t_{F}/t_{T}$ is the ratio of the full transit duration to the total transit duration $t_{T}$, which parametrizes the transit shape.\footnote{The full transit duration is the time between ingress and egress (i.e. second to third contact). The total transit duration is the time between first and fourth contact.} For each planet candidate, we generated posterior probability distributions for the transit observables (i.e. $\delta_{\text{obs}}$, $t_{T}$ and $t_{F}$) by fitting the \TESS{} light curve with  \allesfitter{}  with the transits of the other planet candidates masked out. From these posteriors, we used Eq. \ref{eq:ratio_of_times} to estimate the maximum transit depth caused by the contaminating star ($\delta_{\text{blend}}$) and determine a $3\sigma$ lower limit for $t_{F}/t_{T}$. For all planet candidates, we find $t_{F}/t_{T}\sim0.90$, which suggests that the transits are box-shaped and thus less likely to be caused by a blend \citep{Seager:2003}. 

To determine the \TESS{} magnitude of the faintest blended star ($m_{\text{blend}}$) capable of producing the observed transits in the \TESS{} light curve, we calculated the flux ratio $f$ with Eq. \ref{eq:delta_obs} and exploited the relation between stellar magnitudes and fluxes ($m_{\text{blend}}-m_{\text{star}} =-2.5\log_{10}f$, where  $m_{\text{star}}$ is the \TESS{} magnitude of HD~191939). Our analysis rules out blended stars fainter than $10.13$, $10.38$, and $9.50$ at a $3\sigma$ level for planet candidate b, c, and d, respectively. Therefore, objects such as the spurious \textit{2MASS} detection mentioned in Section \ref{sec:archival_optical_images} would automatically be discarded as the cause of the observed \TESS{} transits. Other nearby stars within the \TESS{} photometric aperture, such as TIC~269701151 (at $42.81$'', with $T=15.63$) and TIC~269701155 (at $45.59$'', with $T = 15.84$), would not be bright enough either to produce the observed transit-like events. Consequently, the results of the ingress-egress test support the planetary nature of the \TESS{} transit signals.

\subsection{Statistical Validation of the HD~191939 System} \label{sec:VESPA_validation}
 
The public software \VESPA{} \citep{VESPA_paper} uses Bayesian inference to calculate the probability that the \TESS{} transits are compatible with astrophysical false positive scenarios. For each HD~191939 planet candidate, we ran \VESPA{} with the planet's \TESS{} phase-folded light curve, the \allesfitter{} best-fit results for its orbital period, transit depth, and planet-to-star radius ratio (see Section \ref{sec:analysis}, Table \ref{tab:MCMC_params}), and the host star's \Gaia{} DR2 coordinates, photometric magnitudes, and effective temperature, metallicity and surface gravity (see Table \ref{tab:TOI_1339_stellar_params}). We also included three observational constraints in our False Positive Probability (FPP) calculation, namely: the Gemini/NIRI contrast curve (see Section \ref{sec:gemini_niri}), a maximum blend radius of $1\arcsec$ based on the high-contrast sensitivity analysis from Section \ref{sec:gemini_niri_validation_test}, and a maximum depth of a potential secondary eclipse of $5\times10^{-5}$. To calculate the latter, we masked out the observed transits of planet candidates b, c, and d on the full \TESS{} discovery light curve, and estimated an upper limit on the shallowest transit depth that could be detected by running a Box-Fitting Least Squares algorithm with the public software \lightkurve{} \citep{Lightkurve:2018}.

The resulting FPPs are less than $10^{-6}$ for all three planet candidates. Given that multiple transit-like signatures are more likely to be caused by genuine planets than by false positives, these FPPs must be enhanced by a "multiplicity boost" corresponding to $\sim15$ for \TESS{} targets (\citealt{Guerrero:2020}). For \TESS{} planets with sizes up to $R_{p}=6\,R_{\oplus}$, this factor increases to $60$. When applying the latter to our \VESPA{} results, we obtain FPPs lower than $10^{-7}$ for all planet candidates. We thus conclude that HD~191939 has three statistically validated \textit{bona fide} planets and refer to them as planet b, c, and d in our subsequent analysis. 

\section{Global Model Fit: Orbital and Planetary Parameters}\label{sec:analysis}

\begin{table*}[!htbp] 
	\centering
	\caption{Final Model Fit Results.}
	\begin{tabular}{lccc}
    	\hline
    	\textbf{Parameter} & HD~191939\,b & HD~191939\,c & HD~191939\,d \\ 
        \hline
        \hline
        Radius Ratio, $R_\mathrm{p} / R_\star$ & $0.03343^{+0.00043}_{-0.00043}$ & $0.03158^{+0.00054}_{-0.00054}$ & $0.03089^{+0.00060}_{-0.00060}$ \\ 
        Sum of Radii over Semi-major axis, $(R_\star + R_\mathrm{p}) / a$ & $0.0553_{-0.0020}^{+0.0023}$ & $0.02548_{-0.00095}^{+0.0011}$ & $0.02084_{-0.00081}^{+0.00097}$  \\ 
        Cosine of Orbital Inclination, $\cos{i}$ & $0.0317^{+0.0036}_{-0.0036}$ & $0.0153_{-0.0016}^{+0.0017}$ & $0.0089^{+0.0020}_{-0.0020}$  \\ 
        Mid-transit Time, $T_{0}$ (BJD days) & $2458715.35554^{+0.00064}_{-0.00064}$ & $2458726.0531_{-0.0011}^{+0.0011}$ & $2458743.5505^{+0.0015}_{-0.0015}$  \\ 
        Orbital Period, $P$ (days) & $8.880403^{+0.000070}_{-0.000070}$ & $28.58059^{+0.00045}_{-0.00045}$ & $38.3561^{+0.0012}_{-0.0012}$   \\ 
        Transit Depth, $\delta_\mathrm{dil; TESS}$ (ppt) & $1.199_{-0.025}^{+0.023}$ & $1.059^{+0.030}_{-0.030}$ & $1.072^{+0.038}_{-0.038}$  \\ 
        Planet Radius, $R_\mathrm{p}$ ($\mathrm{R_{\oplus}}$) & $3.42^{+0.11}_{-0.11}$ & $3.23^{+0.11}_{-0.11}$ & $3.16^{+0.11}_{-0.11}$  \\
        Semi-major Axis, $a$ (AU) & $0.0814^{+0.0040}_{-0.0040}$ & $0.1762^{+0.0089}_{-0.0089}$ & $0.215^{+0.011}_{-0.011}$ \\ 
        Orbital Inclination, $i$ (deg) & $88.18^{+0.21}_{-0.21}$ & $89.124_{-0.097}^{+0.091}$ & $89.49^{+0.12}_{-0.12}$ \\ 
        Impact Parameter, $b_\mathrm{tra}$ & $0.593_{-0.045}^{+0.041}$ & $0.619_{-0.043}^{+0.040}$ & $0.439_{-0.088}^{+0.074}$ \\
        Total Transit Duration, $T_\mathrm{tot}$ (h) & $3.075^{+0.022}_{-0.022}$ & $4.455^{+0.039}_{-0.039}$ & $5.527^{+0.046}_{-0.046}$  \\ 
        Full Transit Duration, $T_\mathrm{full}$ (h) & $2.772^{+0.025}_{-0.025}$ & $4.018^{+0.046}_{-0.046}$ & $5.116_{-0.052}^{+0.056}$ \\ 
        Equilibrium Temperature, $T_\mathrm{eq}$ (K) & $812_{-17}^{+18}$ & $552_{-11}^{+13}$ & $499_{-11}^{+12}$  \\
        \hline \hline
        \multicolumn{4}{c}{\textbf{System Parameters in the \TESS{} band-pass}} \\ 
        \hline
        Limb darkening Coefficient 1, $u_\mathrm{1; TESS}$ & $0.52_{-0.21}^{+0.15}$ & & \\ 
        Limb darkening Coefficient 2, $u_\mathrm{2; TESS}$ & $-0.09_{-0.19}^{+0.27}$ & &\\ 
        Flux error, $\log{\sigma_\mathrm{TESS}}$ ($\log{ \mathrm{rel. flux.} }$) &  $-7.6797^{+0.0028}_{-0.0028}$ & & \\
        GP Characteristic Amplitude, $\mathrm{gp: \log{\sigma} (TESS)}$ & $-9.489^{+0.046}_{-0.046}$ & &  \\
        GP Timescale, $\mathrm{gp: \log{\rho} (TESS)}$ & $-1.32^{+0.13}_{-0.13}$ & & \\ 
        \hline
        \end{tabular}
   \label{tab:MCMC_params}
\end{table*}

We first employed the publicly available Transit Least Squares (\texttt{TLS}) package \citep{TLS} to look for planetary transits in the \TESS{} light curve (see \autoref{fig:lightcurves}).\footnote{The \texttt{TLS} searches for transit-like events in photometric light curves by fitting a physical transit model with ingress, egress, and stellar limb-darkening. This method enhances the detection efficiency by $\sim10\%$ relative to the standard Box Least Squares algorithm (\citealt{Kovacs:2002}), which fits a boxcar function to the transit signatures.} 
The \texttt{TLS} routine identified three possible planetary signals with periods of $P_{b}\approx8.88$\,days, $P_{c}\approx28.58$\,days, and $P_{d}\approx38.35$\,days, confirming the findings of the SPOC pipeline. Taking the \texttt{TLS} orbital periods and transit times as our initial guesses, we performed a preliminary fit to the \TESS{} light curve with \allesfitter{}. For our final fit to the \TESS{} data, we assumed circular orbits and fitted a transit model with nine free parameters: 
\begin{itemize}[noitemsep,topsep=0pt,leftmargin=12pt]
    \item the planet-to-star radius ratio, $R_\mathrm{p}/R_\star$, with uniform priors from 0 to 1,
    \item the sum of the planetary and stellar radii over the semi-major axis, $(R_\mathrm{p} + R_\star) / a$, with uniform priors from 0 to 1,  
    \item the cosine of the orbital inclination, $\cos{i}$, with uniform priors from 0 to 1,  
    \item the planetary orbital period, $P$ with uniform priors from -0.05 to +0.05 days around the initial \texttt{TLS} guess,  
    \item the initial transit time $T_{0}$, with uniform priors from -0.05 to +0.05 days around the initial \texttt{TLS} guess, 
    \item a quadratic stellar limb-darkening function, sampled uniformly with the triangular sampling technique of \citet{Kipping_2013},
    \item white noise (jitter) scaling terms for the \TESS{} light curve, and 
    \item two GP hyper-parameters for the Matern-3/2 kernel: the characteristic amplitude $\ln{\sigma}$, and the timescale $\ln{\rho}$.\vspace{0.2cm}
\end{itemize}

We used a Markov Chain Monte Carlo (MCMC) algorithm, implemented by the Affine-invariant MCMC ensemble sampler \texttt{emcee}, to determine the posterior probability distributions of all the model parameters. We initialized the MCMC with 200 walkers, each taking 40,000 steps, and performed a burn-in of 10,000 steps for each chain before calculating the final posterior distributions. The resulting phase-folded light curves are shown in \autoref{fig:lightcurves} together with the best transit model. The associated fit parameters and their $1\sigma$ uncertainties are listed in Table \ref{tab:MCMC_params}. \autoref{fig:corner_plot} shows the posteriors for these model parameters. To ensure MCMC convergence, we required that the auto-correlation time for each parameter be larger than 30 \citep{emcee}. 

For each planet candidate, we also derived additional system parameters, including  the planetary radius, semi-major axis, orbital inclination, and equilibrium temperature; the stellar density; the ratio of the stellar radius to the planet's semi-major axis; and the light curve observables imprinted by the planet, namely, the diluted transit depth, the total transit duration, the transit duration between ingress and egress, and the impact parameter (see Table \ref{tab:MCMC_params}). 

\section{Discussion}\label{sec:discussion}
\subsection{Dynamical Analysis}

\begin{figure}[!htbp]
  \begin{center}
    \includegraphics[width=\columnwidth]{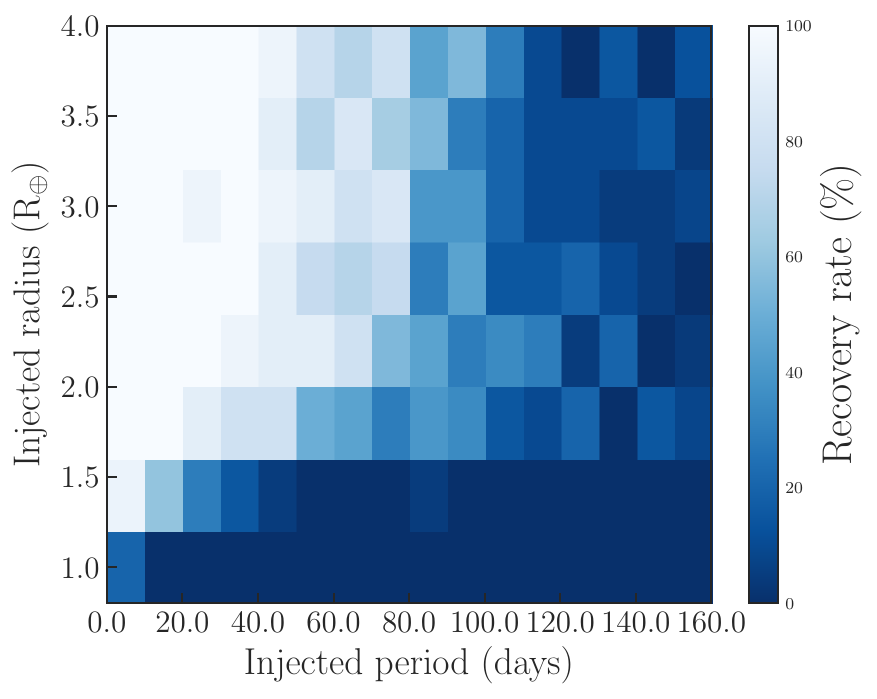}
  \end{center}
  \caption{Injection-recovery test for simulated transits of small planets (0.8 to 4\,\rearth{}; y-axis) on periods between 2 to 160\,days (x-axis). The color-coding shows the completeness of the recovery, with darker tones representing lower recovery rates. The \TESS{} data for HD~191939 collected so far is near-complete for sub-Neptunes and super-Earths on orbits less than $\sim80$\,days, but the regime of the smallest and longest-period planets remains to be explored.}
  \label{fig:injected_transit_search}
\end{figure}

\subsubsection{Orbital Stability} \label{sec:stability}

An important test of orbital architectures derived from observation includes an analysis of the long-term dynamical stability. Such tests have been performed for numerous systems to investigate the validity of Keplerian solutions and the dynamical evolution of the systems (e.g. \citealt{Fabrycky:2014}). For the HD~191939 system, we performed N-body integrations using the Mercury Integrator Package \citep{chambers1999}. Based on the stellar parameters shown in Table \ref{tab:derived_TOI_1339_stellar_params} and the orbital properties listed in Table \ref{tab:MCMC_params}, we constructed a dynamical simulation that spanned $10^7$ simulation years. The starting eccentricity for the planets was fixed to that of circular orbits using the estimated planetary masses from Section \ref{sec:archival_rv_analysis}. Since the innermost planet has a very short orbital period, we chose a conservative time step of 0.1 days to ensure accuracy of the dynamical model. 

The results of the simulation indicate that the HD~191939 system is stable based on the observed orbital parameters. In addition, there is little interaction between the planets and their eccentricities remain below 0.01 for the duration of the simulation, resulting in minimal impact on insolation flux received by the planets that would affect climate \citep{kane2017d}. In particular, the innermost planet retains a circular orbit since it is the most massive and primarily influenced by the host star.

\subsubsection{Potential for Additional Planets}\label{sec:injection_recovery_tests}

In Section~\ref{sec:analysis}, we conducted a full \texttt{TLS} search for a 3\,$\sigma$ detection of a fourth planet and found no evidence of it. To probe our sensitivity limit and investigate the detectability of planets which might still be hidden in the \TESS{} data due to low SNR or data gaps, we also performed a series of injection recovery tests. In particular, we generated planet signals and injected them into the \TESS{} light curves using \allesfitter{}, with planet sizes ranging from 0.8 to 4\,\rearth{} and periods ranging from 2 to 160\,days. We then searched for these signals with \texttt{TLS}, and recorded a signal to be recovered if the detected period matched any multiple of half the injected period to better than 5\%. We find that \TESS{}' recovery is nearly complete for sub-Neptunes and super-Earths on orbits less than $\sim80$\,days days (\autoref{fig:injected_transit_search}). However, the regime of Earth-sized planets and of small exoplanets on longer orbits remain open for exploration. This means that more transiting planets amenable to atmospheric characterization might still await to be discovered in future \TESS{} sectors, while RV and TTV monitoring might unveil additional non-transiting companions. 

\subsubsection{Transit Timing Variations} \label{sec:ttvs}

Transit timing variations (TTVs) in multi-planetary systems are caused by deviations from Keplerian orbits due to gravitational interactions between the planets \citep[e.g.][]{Agol2005, Holman2005, Lithwick2012,Agol2018}. These interactions can be used to verify the planetary nature of a transit signal and to characterize the planetary masses and the system's orbital architecture. TTVs occur for systems in which pairs of planets orbit near mean motion resonances (MMR), where a `first-order MMR' is defined as the period ratio being close to $P_\mathrm{in} / P_\mathrm{out} \approx (i-1)/i$, whereby $P_\mathrm{in}$ and $P_\mathrm{out}$ are the periods of the inner and outer planets, and $i$ is an integer. 
The planets' mid-transit times then show sinusoidal variations with the `TTV super-period', $P_\mathrm{TTV} = {|i/P_\mathrm{out} - (i-1)/P_\mathrm{in}|}^{-1}.$

For HD~191939, planets c and d are close to a first order MMR with a period ratio near 3:4. We would thus expect a TTV super-period of $P_\mathrm{TTV} \sim 1500$\,days -- a factor of 10 longer than the span of our discovery data set ($\sim$150 days). This means we are currently only starting to sample the TTVs of this system, and are still in a regime where the linear period fits for planets c and d are likely biased. In contrast, the inner pair of planets, b and c, lie further off a second-order MMR with a period ratio near 1:3 and are thus expected to show much lower TTV amplitudes. In addition, short-timescale `chopping' variations can occur when the planets are closest to another on their orbits \citep[e.g.][]{Deck2015}. These chopping TTVs typically occur on harmonics of the synodic timescale, 
$P_\mathrm{TTV;chopping} = {|1/P_\mathrm{out} - 1/P_\mathrm{in}|}^{-1}.$ For planets c and d, we expect this to happen on timescales of $\sim 100$\,days, well within the available observation range.

We searched for evidence of TTVs by performing an \allesfitter{} fit to the \TESS{} light curve. For this analysis, we froze the initial epoch and orbital period, and fitted the rest of transit parameters described in Section~\ref{sec:analysis}, with the addition of a TTV parameter for each transit to allow for a shift in the mid-transit time. For independent confirmation, we also used the \texttt{exoplanet} \citep{exoplanet:exoplanet} software and modeled the planetary orbits using the \texttt{TTVOrbit} class with Gaussian priors on the system parameters (from Section~\ref{sec:analysis}, Table~\ref{tab:MCMC_params}). In each study, we placed uniform priors on the mid-transit time of each observed transit in the \TESS{} light curve, centered on the expected mid-transit time from the global fit (Table~\ref{tab:MCMC_params}) with a width of 1 hour. We determined convergence once the fits reached a chain length of at least 30 times the auto-correlation length for \texttt{allesfitter}, and a Gelman-Rubin statistic $\hat{R} < 1.001$ for \texttt{exoplanet} \citep{Gelman1992}. 

\begin{figure}[!htbp]
  \begin{center}  
    \includegraphics[width=0.9\columnwidth]{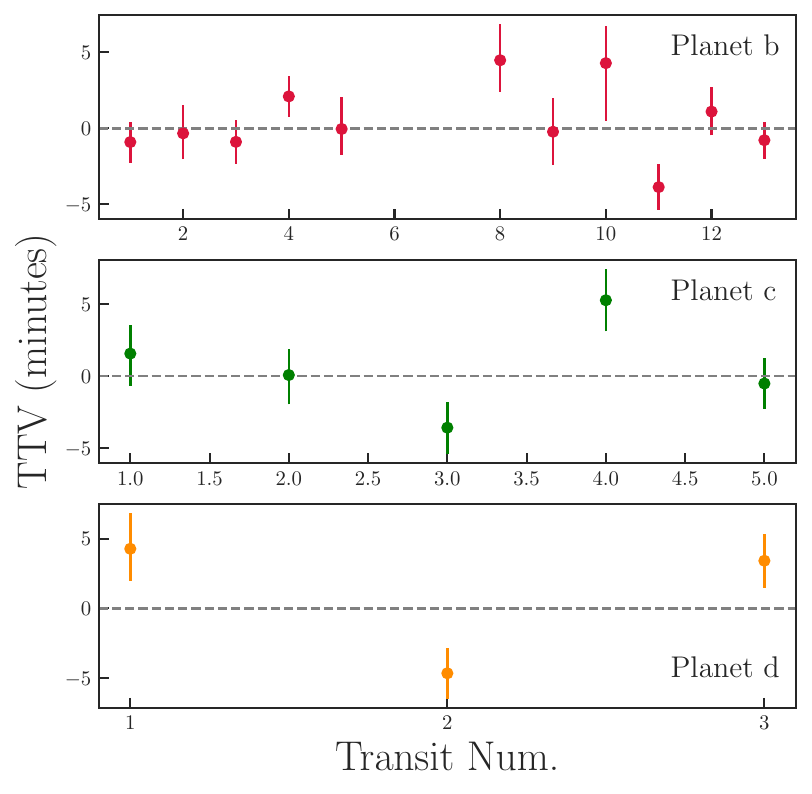}
  \end{center}
  \caption{A search for TTV signals in the \TESS{} data. From a free transit mid-time fit, we find per-transit deviations from linear ephemerides at $\sim2\sigma$, most notably for planets c and d. This could indicate a short-term TTV chopping signal.}
  \label{fig:ttv_results}
\end{figure}

While we find no significant evidence for the long-term super-period TTVs (as expected), we recognize a deviation of the transit midtimes from strictly linear ephemerides on shorter time scales (\autoref{fig:ttv_results}). This could either be due to noise or hint towards a chopping signal. Most notably, the first transit of planet d arrives $4.3 \pm 2.5$\,min. late, the second arrives $4.7 \pm 1.8$\,min. early, and the third arrives $3.4 \pm 1.9$\,min. late again. Similarly, the third transit of planet c arrives $3.6 \pm 1.8$\,min. early and the fourth transit arrives $5.2 \pm 2.1$\,min. late. Future \TESS{} observations and ground-based photometric follow-up will be needed to search for the first conclusive evidence of a chopping signal and to constrain the presence of long-term super-period TTV trends. Initial analyses could be possible after the first full year of monitoring with \TESS{}, when a quarter of the super-period will have been sampled.

\subsection{Atmospheric Characterization Prospects}\label{sec:atmospheric_prospects}

\begin{deluxetable*}{lcccccc}
    \centering
	\tablecaption{\it{Top Panel:} Best confirmed sub-Neptunes ($R_{p}=2-4 R_{\oplus}$) for transmission spectroscopy. We only show planets with measured masses and a relative error in host star radius, planet radius, and planet mass less than $30\%$. Data were retrieved from the NASA Exoplanet Archive in May 2020. \it{Bottom Panel}: Predicted SNR for the HD~191939 planets.}
	\startdata
	\\
    Planet Name             & Host Star & J-magnitude  & Relative SNR$^\star$  &    Planet Radius     & Discovery  & Planet Radius Reference \\ 
                                                 &            & (mag)       &               &    ($R_{\oplus}$)    &            &                            \\
    \hline 
    \noalign{\smallskip}
        1. GJ~436~b         & GJ~436          & 6.90   & $8.35\pm0.14$      & $3.96^{+0.05}_{-0.05}$     & W. M. Keck & \citet{Knutson:2011}   \\
        2. GJ~1214~b        & GJ~1214         & 9.75         & $7.69\pm0.23$      & $2.74^{+0.06}_{-0.05}$     & MEarth     & \citet{Kundurthy:2011} \\
        3. $\pi$~Men~c      & HD~39091        & 4.87   & $4.78\pm0.28$            & $2.06^{+0.03}_{-0.03}$     & \TESS{}       & \citet{Gandolfi:2018}  \\ 
        4. HD~97658~b       & HD~97658        & 6.20         & $2.34\pm0.13$      & $2.247^{+0.098}_{-0.095}$  & W. M. Keck & \citet{VanGrootel:2014} \\
        5. HD~3167~c        & HD~3167         & 7.55  & $1.63\pm0.22$            & $2.85^{+0.24}_{-0.15}$     & K2         & \citet{Vanderburg:2016} \\
        6. GJ~9827~d        & GJ~9827         & 7.98         & $1.45\pm0.32$      & $2.022_{-0.043}^{+0.046}$  & K2         & \citet{Rice:2019}  \\       
        7. TOI-125~c  & TOI-125   & 9.47   & $1.07\pm0.25$      & $2.76_{-0.1}^{+0.1}$       & \TESS{}   & \citet{Nielsen:2020} \\ 
        8. GJ~143~b         & GJ~143          &    6.08       & $1.00\pm0.18$            & $2.61_{-0.16}^{+0.17}$     & \TESS{}       & \citet{Dragomir2019} \\
        9. HD~15337~c & HD~15337  & 7.66   & $0.98\pm0.30$   & $2.55_{-0.10}^{+0.10}$ & \TESS{}  & \citet{Dumusque:2019} \\ 
        10.TOI-125~b  & TOI-125  & 9.47    & $0.91\pm0.18$   & $2.73_{-0.08}^{+0.08}$ & \TESS{}  & \citet{Nielsen:2020} \\
    \hline
    \vspace{1.2pt}
    \textbf{HD~191939~b}\tablenotemark{$^{\dagger}$} & HD~191939 &    7.59     & $1.81\pm0.19$ & $3.42^{+0.11}_{-0.11}$     & \TESS{}       &   This Work       \\ 
    \textbf{HD~191939~c}\tablenotemark{$^{\dagger}$} & HD~191939 &    7.59     &  $1.33^{+0.19}_{-0.20}$ & $3.23^{+0.11}_{-0.11}$     & \TESS{}       & This Work               \\
    \textbf{HD~191939~d}\tablenotemark{$^{\dagger}$}& HD~191939 &    7.59     & $1.29\pm0.20$ & $3.16^{+0.11}_{-0.11}$     & \TESS{}       & This Work             \\
    \enddata
    \tablenotetext{^{\dagger}}{Planetary masses estimated from the MR relation of \citealt{Wolfgang:2016}.}
    \vspace{-1.5pt}
    \tablenotetext{^{\star}}{The predicted SNR for all planets is given relative to that of GJ~143~b.}
    \label{tab:ranking_atmospheric_characterization}
\end{deluxetable*}

\begin{figure*}[!htbp]
  \begin{center}
    \includegraphics[width=0.90\textwidth]{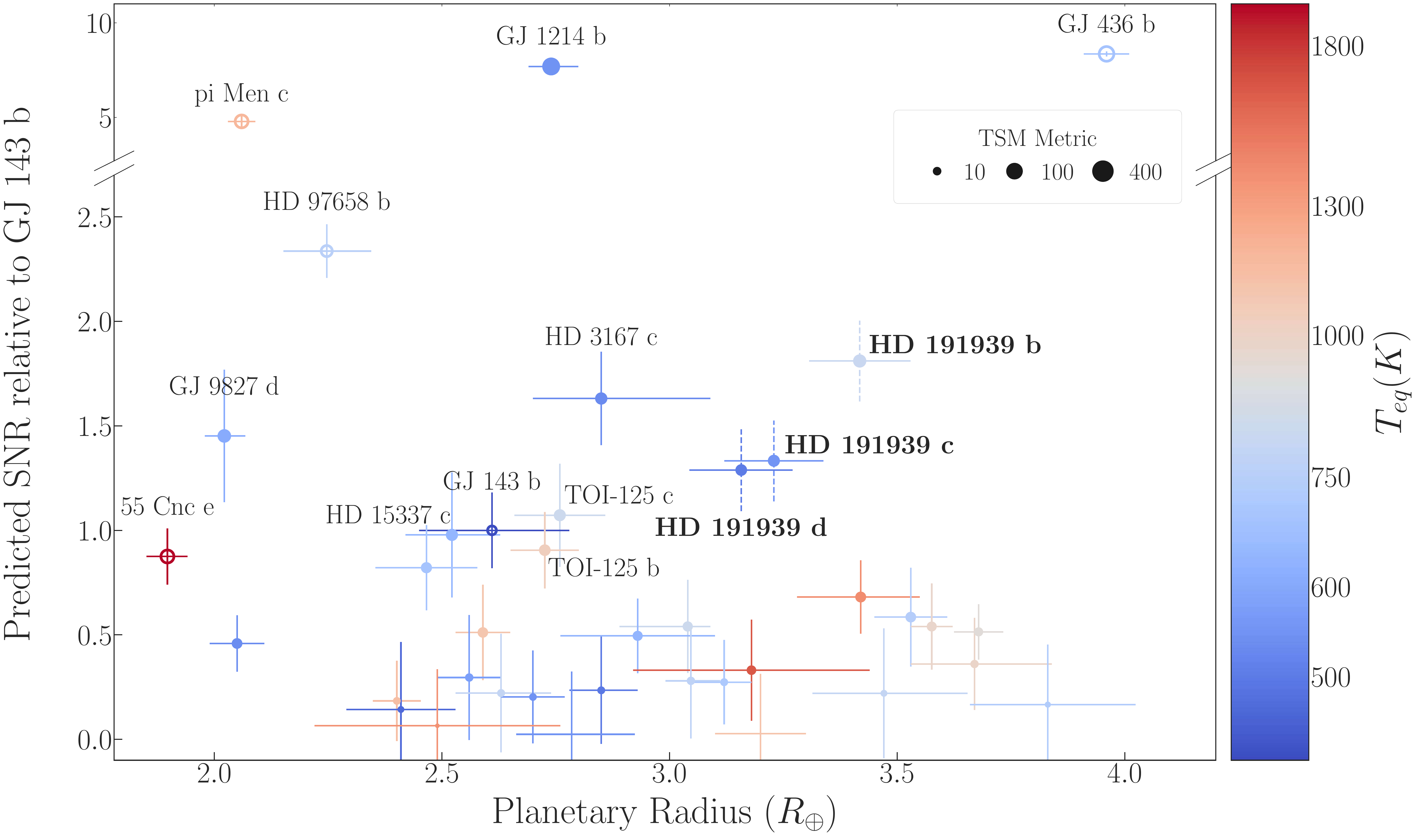}
  \end{center}
  \caption{Predicted SNR relative to GJ~143~b for the population of confirmed sub-Neptunes ($R_{p}=2-4 \,R_{\oplus}$), with the addition of 55~Cnc~e ($R_{p} = 1.897_{-0.046}^{+0.044}\,R_{\oplus}$, \citealt{Dai:2019}) for comparison. We only present systems with mass measurements and a relative error in host star radius, planet radius, and planet mass less than $30\%$. The color of the marker represents the planet's equilibrium temperature, while its size indicates its TSM. Empty circles are used for planets with a host star brighter than the JWST brightness limit ($J\approx7$\,mag; \citealt{Beichman:2014}). Names are only displayed for the top 10 planets (see Table \ref{tab:ranking_atmospheric_characterization}). We show the y-error bars of the HD~191939 planets with dashed lines to indicate that the SNR of the HD 191939 planets is based on mass estimates. Data were retrieved from the NASA Exoplanet Archive on May 2020.}
  \label{fig:atmospheric_prospects}
\end{figure*}

All three HD~191939 planets have the potential to be excellent transmission spectroscopy targets, contingent upon mass measurements. The equilibrium temperatures for a zero Bond albedo and efficient redistribution of heat to the nightside hemisphere are about 800\,K, 550\,K, and 500\,K for planets b, c, and d, respectively. To estimate the amplitude of the atmospheres' transmission signal, we assumed hydrogen-dominated compositions and used the predicted masses from the probabilistic MR relation of \citealt{Wolfgang:2016} ($M_{b}=14.77^{+1.98}_{-1.97},M_{\oplus}$, $M_{c}=13.85
^{+1.87}_{-1.85}\,M_{\oplus}$ and $M_{d}=13.50^{+1.84}_{-1.80}\,M_{\oplus}$). Under these conditions, a change in the planet radius corresponding to one pressure scale height ($H$) would result in a change in the transit depth of $\sim15$\,ppm for HD~191939\,b and $\sim10$\,ppm for both HD~191939\,c and HD~191939\,d. At near-infrared wavelengths, absorption due to species such as H$_{2}$O and CH$_{4}$ can produce variations of a few pressure scale heights in the effective planetary radius, translating to transmission signals $\gtrsim10-80$\,ppm for all three planets. Given the brightness of the host star, this would put all three HD~191939 planets among the most favorable sub-Neptunes currently known for transmission spectroscopy. 

To put HD~191939 in the broader context of confirmed sub-Neptunes suitable for atmospheric characterization, we downloaded a list from the NASA Exoplanet Archive of all the confirmed planets with radii between 2--4\,$R_{\oplus}$. We also required these planets to have mass measurements and a relative error in planet mass, planet radius, and host star radius less than $30\%$. For planets with multiple measurements of a given parameter, we selected the reported value with the lowest total uncertainty. We then calculated the expected SNR of each planet for a single transit as \citep{Vanderburg:2016}: 
\begin{equation}\label{eq:SNR}
    SNR \propto \frac{R_{p}H\sqrt{Ft_{14}}}{R^{2}_{*}},
\end{equation}
with the atmosphere's scale height given by $H = \frac{k_{b}T_{\text{eq}}}{\mu g}.$
Above, $F$ is the stellar flux, $k_{b}$ is the Boltzmann's constant, $T_{\text{eq}}$ is the planet's equilibrium temperature, $\mu$ is the atmospheric mean molecular weight, $g$ is the planet's surface gravity, and $t_{14}$ is the transit duration \citep{seager2010_exoplanetatmospheres}. We computed the stellar flux from the host star's H-band \textit{2MASS} magnitude and set $\mu=4$ atomic mass units (amu), corresponding to approximately 100x solar metallicity, in line with sub-Neptune formation simulations \citep{Fortney:2013}. In the absence of $t_{14}$ and $T_{\text{eq}}$ values listed on the Exoplanet Archive, we computed these parameters with Eq. 16 and Eq. 2.27 in \citealt{Seager:2003} and \citealt{seager2010_exoplanetatmospheres}, respectively. For these calculations, we assumed a zero Bond albedo ($A_{B}=0$) and full heat redistribution over the planet's surface ($f' = 1/4$). 

Table \ref{tab:ranking_atmospheric_characterization} shows the HD~191939 planets and the best known sub-Neptunes for atmospheric characterization work, according to Eq. \ref{eq:SNR}. Our study indicates that all three HD~191939 planets may be valuable candidates for transmission spectroscopy, with HD~191939\,b offering the highest SNR predictions, followed by HD~191939\,c and HD~191939\,d. However, it is necessary to measure the planetary masses before this can be confirmed. Moreover, the relatively long periods of the HD~191939~b and c planets ($\sim$29 and 38 days; see Table \ref{tab:MCMC_params}) may limit the number of transit events per observing campaign. This may make it more challenging to schedule and obtain the necessary observations to build up the required SNR for atmospheric characterization work, especially in comparison with the shorter-period planets in Table \ref{tab:ranking_atmospheric_characterization}, such as GJ~436~b ($P\approx2.6$~days) or GJ~1214~b ($P\approx1.6$~days). 

\autoref{fig:atmospheric_prospects} shows the HD~191939 planets in the context of the sub-Neptunes considered in this study. It also illustrates the planets' equilibrium temperatures and their Transmission Spectroscopy Metric (TSM; \citealt{Kempton:2018}).\footnote{The TSM predicts the expected transmisison spectroscopy SNR of a 10-hour observing campaign with JWST/NIRISS under the assumption of a fixed MR relationship, cloud-free atmospheres, and the same atmospheric composition for all planets of a given type.} From an anticipated SNR perspective, HD~3167\,c \citep{Vanderburg:2016} offers a useful point of comparison. The latter is a 2.9$R_{\oplus}$ planet orbiting a bright ($J=7.5$\,mag) K0 V host star, with a period of about 30 days, an equilibrium temperature of $600$\,K and a transmission signal amplitude of $\sim20$\,ppm for a $1H$ change in effective planet radius. Given the similar brightness of the HD~3167 and HD~191939 host stars, this means that HD~191939\,c and HD~191939\,d could be comparably suited for transmission spectroscopy, and that HD~191939\,b could be more favorable than HD~3167\,c, owing to its higher equilibrium temperature and thus larger atmospheric scale height.

\subsection{HD~191939 in Context}

\begin{figure*}[htbp]
    \centering
    \includegraphics[width=0.85\linewidth]{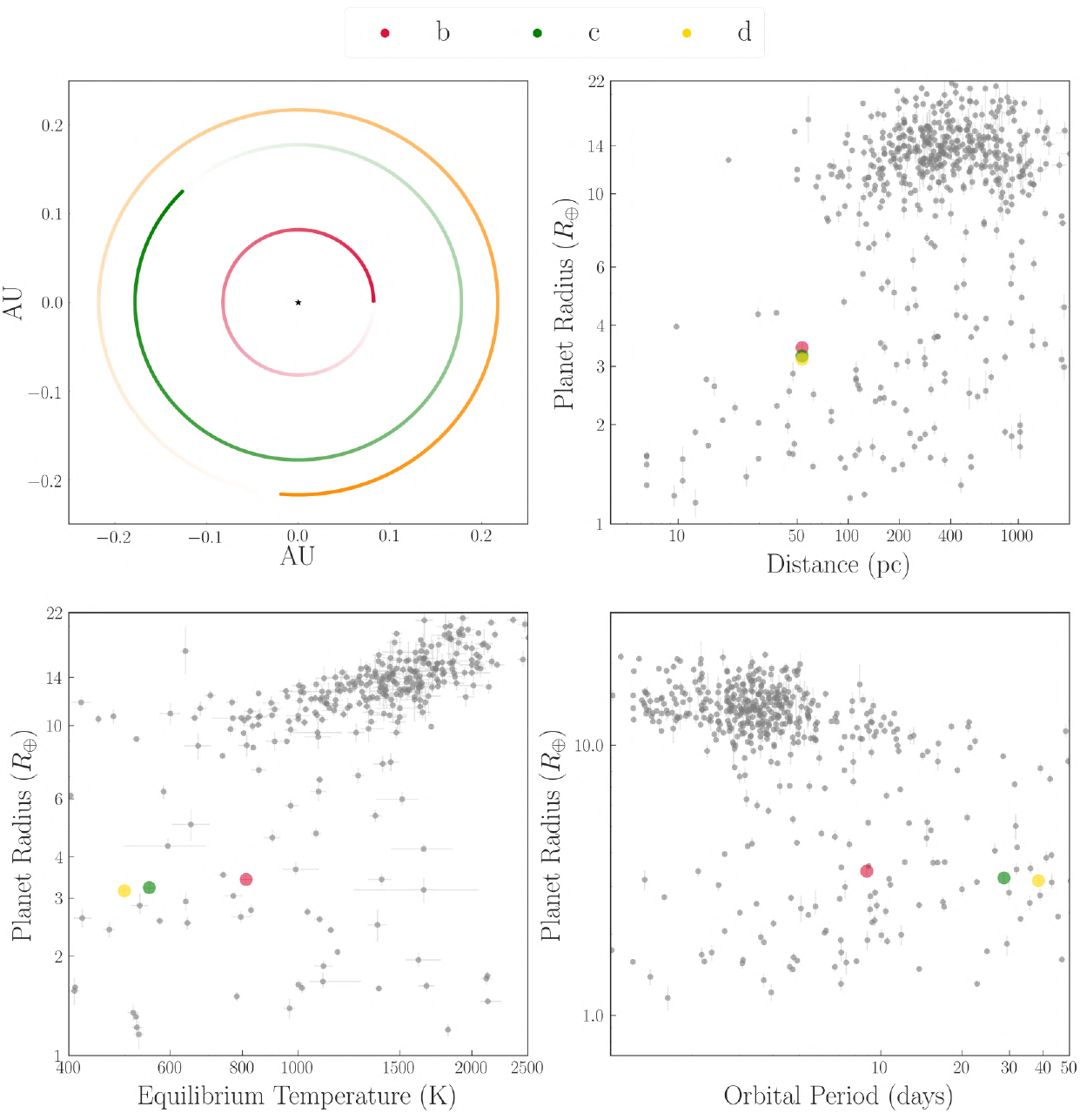}
    \caption{
    \label{fig:toi1339_planets}
    Planets b, c and d are shown in red, green, and orange, respectively. \textit{Top left}: Top-down view of the HD~191939 system, with the sizes of the planets drawn to scale. \textit{Top right, bottom left, and bottom right}: Planet radius as a function of distance, equilibrium temperature, and orbital period, respectively. The systems shown in these views have mass measurements for at least one of their planets and a relative error in host star radius, planet radius, and planet mass better than $30\%$. Data were retrieved from the NASA Exoplanet Archive on May 2020.}
\end{figure*}

The Sun-like star HD~191939 hosts three transiting sub-Neptunes in a compact orbital configuration. This system is a promising candidate for detailed characterization, as evidenced by \autoref{fig:rs_kmag_vs_dist} and \autoref{fig:toi1339_planets}.

\begin{deluxetable*}{lcccc}
    \centering
	\tablecaption{\label{tab:stellar_hosts} HD~191939 relative to confirmed multi-planetary systems with FGK stellar hosts located at a distance of less than 100\,pc (sorted by K-magnitude). Data were retrieved from the NASA Exoplanet Archive in May 2020.}
	\startdata
	\\
    Host Name    & K$_{s}$-magnitude & Distance  & Spectral Type   &  Known Planets \\
                 &     (mag)         &   (pc)    &                 &     (Num.)      \\
    \hline 
    \noalign{\smallskip}
    HD~219134           &  3.26     &    6.53    &     K3 V         &        6      \\
    55~Cnc              &  4.02     &   12.59    &     G8 V         &        5      \\
    HD~39091            &  4.24 	&   18.27    &     G0 V         &        2      \\
    GJ~143	            &  5.38     &   16.32    &     K4.5         &        2      \\
    HD~213885           &  6.42     &	48.09    &      G           &        2      \\
    HAT-P-11	        &  7.01     &   37.76	 &      K4          &        2      \\
    HD~15337        	&  7.04     &	44.81    &      K1 V        &        2      \\
    HD~3167             &  7.07     & 	47.29    &     K0 V         &        3      \\
    \textbf{HD~191939}  &  7.18     &   53.48    &     G8 V         &        3      \\
    GJ~9827             &  7.19	    &   29.66    &     K5 V	        &        3	    \\
    WASP-8              &  8.09  	&   89.96	 &      G8 V        &        2      \\
    TOI-1130        	&  8.35	    &   58.26	 &      K           &        2      \\
    Kepler-93       	&  8.37	    &   95.91    &      G5           &        2      \\	
    K2-141	            &  8.40	    &   61.87	 &      K7 V        &        2      \\
    HAT-P-17            &  8.54	    &   92.38	 &      K           &        2      \\
    \enddata
\end{deluxetable*}

First, the host star's brightness and proximity make HD~191939 an excellent target for future photometric follow-up. In the context of single- and multi-planetary systems for which mass measurements are available and the relative error in host star radius, planet mass, and planet radius is less than $30\%$ (see \autoref{fig:rs_kmag_vs_dist}), HD~191939 is one of the brightest and closest multis known to date. This also holds true when comparing HD~191939 to nearby ($d\leq100$\,pc) multi-planetary systems with Sun-like (FGK) stellar hosts (see Table \ref{tab:stellar_hosts}).

A closer look at the properties of the systems listed in Table \ref{tab:stellar_hosts} reveals several interesting connections between HD~191939 and GJ~9827 \citep{Niraula:2017}. The latter is composed of two planets in the super-Earth regime (GJ~9827~b and c) and an outer mini-Neptune (GJ~9827~d). Similarly to HD~191939, GJ~9827 is a triplet in which all three planets transit their bright parent star. Both systems are near mean-motion resonances, with the former presenting a possible first and second order MMR (see Section \ref{sec:ttvs}), and the latter featuring period ratios near commensurability of 1:3:5. Finally, both HD~191939 and GJ~9827 exhibit similar SNR predictions for transmission spectroscopy (see Table \ref{tab:ranking_atmospheric_characterization}). More specifically, GJ~9827 offers favorable prospects for the atmospheric characterization of its outer sub-Neptune. In the case of HD~191939, all three sub-Neptunes may be suitable for transmission spectroscopy. Such a study would offer the opportunity to perform a comparative study of the planets' atmospheres and investigate the fundamental properties of multiple sub-Neptunes born with a similar formation and evolutionary history. 

HD~191939 is also a valuable target for follow-up photometry due to its location in the northern ecliptic hemisphere sky. Indeed, HD~191939 lies in a region where 10 \TESS{} sectors overlap (Sectors 14--19, 21--22, 24--25), hence enabling a long \TESS{} observational baseline. In particular, HD~191939 will have been monitored for about 10 months once \TESS{} Sector 25 (2020 June 08) concludes and will be re-observed for an additional $\sim$ 10 months during the \TESS{} extended mission. As a result, HD~191939 will have a photometric baseline of almost $\sim$3 years. Such a long time span will facilitate a variety of dynamical studies, including a refinement of the system's transit ephemerides and a search for additional planetary companions via TTV analyses. Moreover, the proximity of the host star ($\sim54$\,pc) will also facilitate searches for massive planetary companions on wide orbits ($\sim$1\,AU) using \Gaia{} astrometry (\citealt{Perryman:2014}).  
 
From the perspective of ground-based RV follow-up, HD~191939 is also an excellent target for future observations due to its brightness, slow rotation, and lack of measurable chromospheric activity. Future RV monitoring, with spectrographs such as CARMENES \citep{Quirrenbach:2014, Quirrenbach:2018} or HARPS \citep{Cosentino:2012_1VC}, will soon enable precise measurements of the planets' masses. A RV monitoring campaign carried out by the CARMENES consortium is currently ongoing to confirm and further characterize the HD~191939 planets. Due to the system's complex orbital architecture, however, a large number of measurements will be needed to accurately constrain the physical properties of the system. 

With three temperate sub-Neptunes, HD~191939 may also be a prime system for atmospheric studies with present (e.g. the \textit{Hubble Space Telescope}) and future (e.g. JWST, Extreme Large Telescopes) facilities. With a high SNR for multi-wavelength transmission spectroscopy (see Table \ref{tab:ranking_atmospheric_characterization}), the three HD~191939 planets appear to be among the best candidates for atmospheric characterization work (\autoref{fig:atmospheric_prospects}). Their final suitability will be confirmed when mass measurements become available. With regard to JWST in particular, HD~191939 will be visible for more than 200 days per year due to its high ecliptic latitude (see Table \ref{tab:TOI_1339_stellar_params}).\footnote{\url{https://jwst-docs.stsci.edu/jwst-observatory-hardware/jwst-target-viewing-constraints}} Future observations will provide an opportunity to study the atmospheres, interiors, and habitability conditions of the HD~191939 planets. 

Finally, the multi-planetary nature of HD~191939 provides a fantastic opportunity to perform comparative exoplanetology. By studying the physical and orbital characteristics of HD~191939 and comparing them to the properties of the known population of multi-transiting planetary systems, we may gain insight into the distribution and occurrence rates of planets across a wide range of masses and radii, as well as into the formation and evolution of planetary architectures.
  
\section{Conclusion} \label{sec:conclusion}

We have presented the \TESS{} discovery of three sub-Neptune-sized planets around the nearby, bright Sun-like star HD~191939. We have confirmed the planetary nature of each planet candidate, both empirically through optical, photometric, and spectroscopic observations, and statistically via the public FPP implementation \VESPA{}. Upon refining the stellar parameters of HD~191939 reported by TICv8, we have derived the physical and orbital properties of the system with an \allesfitter{} fit to the \TESS{} discovery light curve. In addition, we have conducted a dynamical study of the HD~191939 planets, which indicates they are in a compact and stable orbital configuration consistent with circular orbits. Finally, we have demonstrated that the system is a promising target for precise photometric and RV follow-up as well as for future atmospheric characterization studies. 

\section*{Acknowledgments}
Funding for this research is provided by the Massachusetts Institute of Technology, the MIT Torres Fellow Program, and the MIT Kavli Institute.
%TESS Alerts
We acknowledge the use of public \TESS{} Alert data from pipelines at the \TESS{} Science Office and at the \TESS{} Science Processing Operations Center. 
%TESS Funding
Funding for the \TESS{} mission is provided by NASA’s Science Mission directorate. 
%TESS ExoFOP
This research has made use of the Exoplanet Follow-up Observation Program website, which is operated by the California Institute of Technology, under contract with the National Aeronautics and Space Administration under the Exoplanet Exploration Program.
%TESS NASA
Resources supporting this work were provided by the NASA High-End Computing (HEC) Program through the NASA Advanced Supercomputing (NAS) Division at Ames Research Center for the production of the SPOC data products.
%TESS MAST
This paper includes data collected by the \TESS{} mission, which are publicly available from the Mikulski Archive for Space Telescopes (MAST). STScI is operated by the Association of Universities for Research in Astronomy, Inc. under NASA contract NAS 5-26555.
% NASA Exoplanet Archive. 
This research has made use of the NASA Exoplanet Archive, which is operated by the California Institute of Technology, under contract with the National Aeronautics and Space Administration under the Exoplanet Exploration Program. 
% PALOMAR I and II + Pan-Starrs
The Digitized Sky Surveys were produced at the Space Telescope Science Institute under U.S. Government grant NAG W-2166. The images of these surveys are based on photographic data obtained using the Oschin Schmidt Telescope on Palomar Mountain and the UK Schmidt Telescope. The plates were processed into the present compressed digital form with the permission of these institutions. The National Geographic Society - Palomar Observatory Sky Atlas (POSS-I) was made by the California Institute of Technology with grants from the National Geographic Society. The Second Palomar Observatory Sky Survey (POSS-II) was made by the California Institute of Technology with funds from the National Science Foundation, the National Geographic Society, the Sloan Foundation, the Samuel Oschin Foundation, and the Eastman Kodak Corporation.
%Gaia
This work has made use of data from the European Space Agency (ESA) mission Gaia (\url{https://www.cosmos.esa.int/ gaia}), processed by the Gaia Data Processing and Analysis Consortium (DPAC, \url{https://www.cosmos.esa.int/web/gaia/ dpac/consortium}). Funding for the DPAC has been provided by national institutions, in particular the institutions participating in the Gaia Multilateral Agreement. 
% SOPHIE
This paper used data retrieved from the \SOPHIE{} archive at Observatoire de Haute-Provence (OHP), available at \url{atlas.obs-hp.fr/sophie}.
%Gemini
The AO images presented in this paper were obtained at the Gemini Observatory (Program ID: GN-2019B-LP-101), which is operated by the Association of Universities for Research in Astronomy, Inc., under a cooperative agreement with the NSF on behalf of the Gemini partnership: the National Science Foundation (United States), National Research Council (Canada), CONICYT (Chile), Ministerio de Ciencia, Tecnolog\'{i}a e Innovaci\'{o}n Productiva (Argentina), Minist\'{e}rio da Ci\^{e}ncia, Tecnologia e Inova\c{c}\~{a}o (Brazil), and Korea Astronomy and Space Science Institute (Republic of Korea).
%Photometric follow-up attempts

The authors thank Amanda Bosh (MIT), Tim Brothers (MIT Wallace Astrophysical Observatory), Julien de Wit (MIT), Artem Burdanov (MIT), Songhu Wang (Yale University), Enrique Herrero (IEEC/OAdM), Jonathan Irwin (Harvard-CfA), Samuel Hadden (Harvard-CfA), Özgür Baştürk (Ankara University), Ergün Ege (Istanbul University), and Brice-Olivier Demory (University of Bern) and for helping to coordinate follow-up observations. 

%People
We thank the anonymous referee for their helpful comments and suggestions, which greatly improved this work.
M.N.G. and C.X.H. acknowledge support from MIT's Kavli Institute as Juan Carlos Torres Fellows.
T.D. acknowledges support from MIT’s Kavli Institute as a Kavli postdoctoral fellow.
A.V.'s work was performed under contract with the California Institute of Technology/Jet Propulsion Laboratory funded by NASA through the Sagan Fellowship Program executed by the NASA Exoplanet Science Institute.
I.R. acknowledges support from the Spanish Ministry of Science, Innovation and Universities (MCIU) and the Fondo Europeo de Desarrollo Regional (FEDER) through grant PGC2018-098153-B-C33, as well as the support of the Generalitat de Catalunya/CERCA program. 
B.V.R. and J.N.W. thank the Heising-Simons foundation for support. 
I.J.M.C. acknowledges support from the NSF through grant AST-1824644, and from NASA through Caltech/JPL grant RSA-1610091.  

% Facilities
\textbf{Facilities:} TESS, FLWO: 1.5m (TRES), LCO: 1m (NRES), OHP: 1.93m (SOPHIE), Gemini/NIRI, OAA: 0.4m. 

% Software
\textbf{Software:} \texttt{Python} (\citealt{Python}), \texttt{numpy} (\citealt{numpy}), \texttt{scipy} (\citealt{scipy}), \linebreak 
\texttt{matplotlib} (\citealt{matplotlib}), \texttt{astropy} (\citealt{astropy:2018}), \texttt{pandas} (\citealt{pandas:2010}), \allesfitter{} (\citealt[][and in prep.]{allesfitter}) \texttt{emcee} (\citealt{emcee}), \texttt{corner} (\citealt{corner}), \texttt{tqdm} (doi:10.5281/zenodo.1468033), \texttt{lightkurve} (\citealt{Lightkurve:2018}), \texttt{Transit Least Squares} (\citealt{TLS}), \VESPA{} (\citealt{VESPA_paper}), \isochrones{} (\citealt{Morton:2015}), \isoclassify{} (\citealt{Huber:2017}), \forecaster{} (\citealt{chen2017}), \texttt{exoplanet} \citep{exoplanet:exoplanet}, \texttt{starry} \citep{exoplanet:luger18}, \texttt{pymc3} \citep{exoplanet:pymc3}, \texttt{theano} \citep{exoplanet:theano}, \texttt{rebound} \citep{rebound:2012}.

\newpage
\bibliography{references}
\bibliographystyle{aasjournal}

\newpage
\appendix

\begin{figure*}[!htbp]
    \centering
    \includegraphics[width=1\textwidth]{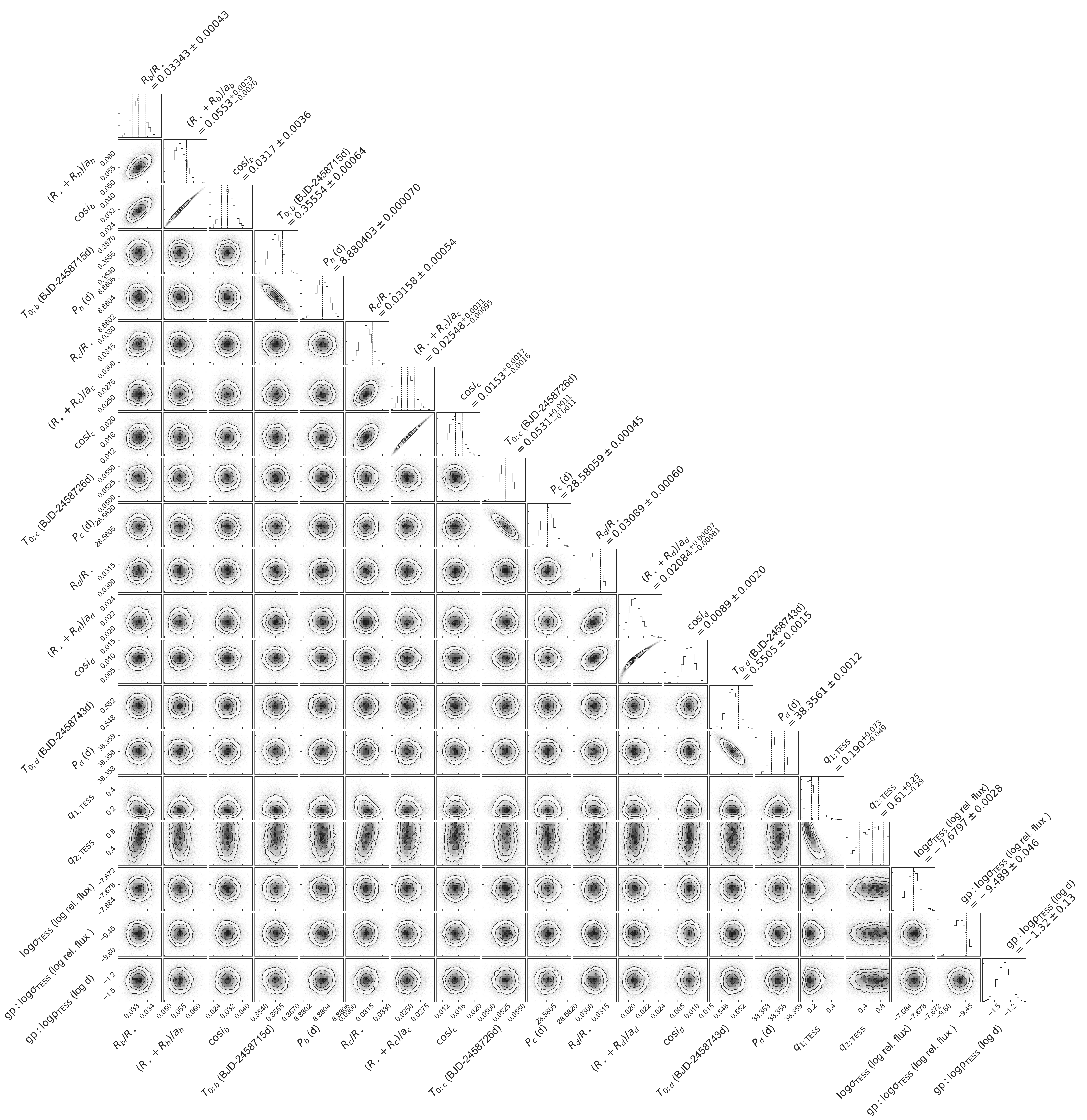}
    \caption{Posterior probability distributions for the \allesfitter{} model parameters. The dashed lines show the 16th, 50th, and 84th percentiles.}
    \label{fig:corner_plot}
\end{figure*}

\end{document}